\documentclass[twocolumn]{aastex631}
\usepackage{amsmath}
\usepackage{enumitem}

\newcolumntype{?}{!{\vrule width 2pt}}

\newcommand{\Teq}{T$_{\rm{eq}}$}
\newcommand{\Teff}{T$_{\rm{eff}}$}
\newcommand{\Rearth}{R$_{\rm{\oplus}}$}
\shorttitle{Vetting a Representative Exoatmospheric Target List}
\shortauthors{Burt \& Zellem et al. 2025}
\graphicspath{{./}{figures/}}

\begin{document}

\title{A New Approach to Compiling Exoatmospheric Target Lists And Quantifying the Ground-Based Resources Needed to Vet Them}

\author[0000-0002-0040-6815]{Jennifer A. Burt}
\affiliation{Jet Propulsion Laboratory, California Institute of Technology, 4800 Oak Grove Drive, Pasadena, CA 91109, USA}

\author[0000-0012-3245-1234]{Robert T. Zellem}
\affiliation{Goddard Space Flight Center, 8800 Greenbelt Rd, Greenbelt, MD 20771, USA}

\author[0000-0002-5741-3047]{David R. Ciardi}
\affiliation{NASA Exoplanet Science Institute-Caltech/IPAC, 1200 East California Boulevard, Pasadena, CA 91125, USA}

\author[0000-0001-8401-4300]{Shubham Kanodia}
\affiliation{Earth and Planets Laboratory, Carnegie Science, 5241 Broad Branch Road, NW, Washington, DC 20015, USA}

\author[0000-0001-5966-837X]{Geoffrey Bryden}
\affiliation{Jet Propulsion Laboratory, California Institute of Technology, 4800 Oak Grove Drive, Pasadena, CA 91109, USA}

\author[0000-0003-3759-9080]{Tiffany Kataria}
\affiliation{Jet Propulsion Laboratory, California Institute of Technology, 4800 Oak Grove Drive, Pasadena, CA 91109, USA}

\author[0000-0002-5785-9073]{Kyle A. Pearson}
\affiliation{Jet Propulsion Laboratory, California Institute of Technology, 4800 Oak Grove Drive, Pasadena, CA 91109, USA}

\author[0000-0002-8035-4778]{Jessie L. Christiansen}
\affiliation{NASA Exoplanet Science Institute-Caltech/IPAC, 1200 East California Boulevard, Pasadena, CA 91125, USA}

\author[0000-0002-5627-5471]{Charles Beichman}
\affiliation{Jet Propulsion Laboratory, California Institute of Technology, 4800 Oak Grove Drive, Pasadena, CA 91109, USA}
\affiliation{NASA Exoplanet Science Institute-Caltech/IPAC, 1200 East California Boulevard, Pasadena, CA 91125, USA}

\author[0000-0003-3504-5316]{B.J. Fulton}
\affiliation{NASA Exoplanet Science Institute-Caltech/IPAC, 1200 East California Boulevard, Pasadena, CA 91125, USA}

\author[0009-0001-4487-7299]{Mark Swain}
\affiliation{Jet Propulsion Laboratory, California Institute of Technology, 4800 Oak Grove Drive, Pasadena, CA 91109, USA}

\begin{abstract}

Transiting exoplanet atmospheric characterization is currently in a golden age as dozens of exoplanet atmospheres are being studied by NASA's Hubble and James Webb Space Telescopes. This trend is expected to continue with NASA's Pandora Smallsat and Roman Space Telescope and ESA's Ariel mission (all expected to launch within this decade) and NASA's Habitable Worlds Observatory (expected to launch in the early 2040s) all of which are centered around studying the atmospheres of exoplanets. Here we explore a new approach to constructing large scale exoatmospheric survey lists, which combines the use of traditional transmission/emission spectroscopy figures of merit with a focus on more-evenly sampling planets across a range of radii and equilibrium temperatures. After assembling a sample target list comprised of 750 transmission spectroscopy targets and 150 emission spectroscopy targets, we quantify the potential time lost to stale transit and eclipse ephemerides and find that hundreds of hours of space-based observing could be wasted given current uncertainties in orbital periods, transit epochs, and orbital eccentricities. We further estimate the amount of ground-based telescope time necessary to obtain sufficiently precise exoplanet masses and find that it exceeds 100 nights of 10m telescope time. Based upon these findings, we provide a list of recommendations that would make community efforts for preparation and interpretation of atmospheric characterization endeavors more effective and efficient. The strategies we recommend here can be used to support both current (e.g., HST and JWST) and future exoplanet atmosphere characterization missions (e.g., Pandora, Ariel, Roman, and the Habitable Worlds Observatory).

\end{abstract}


\section{Introduction} \label{sec:intro}

Over the past thirty years, scientists have moved from wondering if planets exist around other stars to performing statistical studies on the physical characteristics of numerous exoplanet populations. The detection of thousands of transiting exoplanets across a range of masses and sizes, from programs like the ground-based HATNet \citep{Hartman2011} and WASP \citep{Butters2010} surveys and the space-based \textit{Kepler} \citep{Borucki2010} and \textit{TESS} \citep{Ricker2015} missions, has opened up the possibility of atmospheric characterization across a breadth of planetary properties. A remarkably vast discovery space has been enabled by spectroscopic observations during an exoplanet's transit or eclipse that provide information about its atmospheric properties. Spectroscopic facilities both on the ground and in space, including CRIRES+ \citep[e.g.,][]{Dorn2023}, CARMENES \citep[e.g.,][]{Quirrenbach2014}, \textit{Spitzer} \citep[e.g.,][]{Werner2004}, and \textit{Hubble} \citep[e.g.,][]{Sing2011} have been utilized to understand exoplanetary atmospheric compositions and structures, including the detection of individual atomic and molecular species as well as evidence for clouds and hazes. Exoplanet atmospheric science is currently undergoing a period of rapid advancement thanks to the advent of \textit{JWST} \citep[e.g.,][]{Greene2016} which has already observed over 200 exoplanets in the near/mid-IR \citep[see][and references therein]{Espinoza2025}.

The next paradigm shift in our understanding of exoplanetary atmospheres will come with the upcoming launches of NASA's Pandora mission \citep[2025;][]{quintana21}, ESA's Ariel mission \citep[2029;][]{tinetti16, tinetti18, zellem19, edwards19}, and NASA's Habitable Worlds Observatory ($\sim$2040). ESA's Ariel mission aims to conduct a uniform survey of $\sim$1000 exoplanets in transmission spectroscopy and $\sim$200 in emission spectroscopy to enable a statistical assessment of their atmospheres and presents a particularly interesting opportunity. It will, for the first time, provide the exoplanet community with a statistically-significant population of planetary atmospheres with which to form the basis of a new generation of comparative planetology studies that can consider the effects of planet mass, atmospheric composition, metallicity, and equilibrium temperature \citep[e.g.,][]{fortney13, crossfieldkreidberg17, welbanks19}.

To accurately interpret the results of these missions and determine the chemical composition of an exoplanet's atmosphere, the planet's surface gravity must be precisely known \citep[see, e.g.,][]{batalha2018}. The surface gravity is calculated from the planet's observed mass and radius, and so these properties must in turn be well characterized. The radii of transiting exoplanets are derived from their transit depths, while exoplanet masses must be derived either from radial velocity (RV) measurements \citep{Mayor1995, Cumming2004} or transit timing variations in near-resonant, multi-planet systems \citep{Lithwick2012}. 

The majority of transiting exoplanet targets, however, lack any level of mass estimate. And many do not have sufficiently well constrained orbital parameters (e.g., transit epoch, orbital period, eccentricity, argument of periastron, etc.) to ensure precise knowledge of their orbital ephemerides. Indeed the uncertainties on the future times of transit and eclipse are already large enough to impact scheduling of HST and JWST \citep[e.g.,][]{zellem20, dragomir20, pearson22} and will continue to grow year after year such that many hours per target could be wasted or the transits/eclipses could be missed altogether \citep{zellem20}.

In light of these challenges, we put forth the idea to compile a large, representative, exoatmospheric target list of transiting exoplanets and then quantify the scope of a ground-based science program featuring both transit and radial velocity facilities that could provide high precision ephemerides, masses, and orbital parameters for the selected planets. Given the expected scope of the JWST and Ariel prime mission exoplanet atmosphere science programs (100s to 1000s of planets) we demonstrate that such dedicated ground-based programs will exceed the limited telescope resources and observing time available in a given semester, year, or even multi-year campaign. Thus, such a program could benefit from an understanding of which targets are most likely to be observed by future transiting exoplanet observing programs, which targets are most in need of a mass determination, orbital parameter refinement, or ephemeris maintenance, and how much ground-based telescope time is needed to provide these measurements. Our creation of a potential target list and quantification of the ground-based campaigns needed to support it are relevant to any current or future transiting exoplanet missions, such as Hubble, JWST, Pandora, and the Habitable Worlds Observatory.

We note, however, that the target lists generated herein are not meant to serve as definitive lists for near-term vetting and characterization efforts. Rather this work lays out a general methodology for assembling target lists that prioritize a more even distribution in planet size and equilibrium temperature. We emphasize that the set of input planets / planet candidates, the relevant axes to bin those objects along, and the figure of merit applied to rank the objects within those bins should be all revisited and selected in response to the specific science case(s) that are of most interest to a given mission or survey. See, e.g., \citealt{Batalha2023} for an example of defining a figure of merit and binning combination optimized for particular exoatmospheric science questions and \citealt{Cowan2025} for a survey design approach that considers the host stars' effective temperatures as a third axis when creating target bins.

We begin in Section \ref{sec:TargetLists} by building a list of viable transit and eclipse spectroscopy targets, compiling known planets and planet candidates from the NASA Exoplanet Archive, the TESS Objects of Interest list and the Kepler Objects of Interest list and applying a new mass-radius relation to estimate their masses. In Section \ref{sec:RepTargetList} we develop a novel survey target list that combines a traditional figure of merit that considers the expected strength of their spectral modulation, for transits, or eclipse depth, for eclipses, and host star brightness but requires that the list be smoothly populated across a planet radius versus equilibrium temperature phase-space. In Section \ref{sec:TransitUncert} we quantify the cumulative uncertainty of the mid-transit and mid-eclipse times, the impact of the on-going TESS mission, and the amount of ground-based photometry necessary to refresh the planets' transit ephemerides. Then in Section \ref{sec:RVTime} we compute the RV facility time needed to obtain planet masses at the necessary precision based upon each planet's radius and equilibrium temperature and intended atmospheric characterization approach. Section \ref{sec:discussion} discusses the observational time required and ways to prioritize the targets within each observing effort, and then we conclude in Section \ref{sec:conclusions} with a set of recommendations that could make community preparation efforts more effective and efficient.

\section{Establishing a List of Viable Atmospheric Characterization Targets}\label{sec:TargetLists}

\begin{figure*}[htb!]
\includegraphics[width=.98\textwidth]{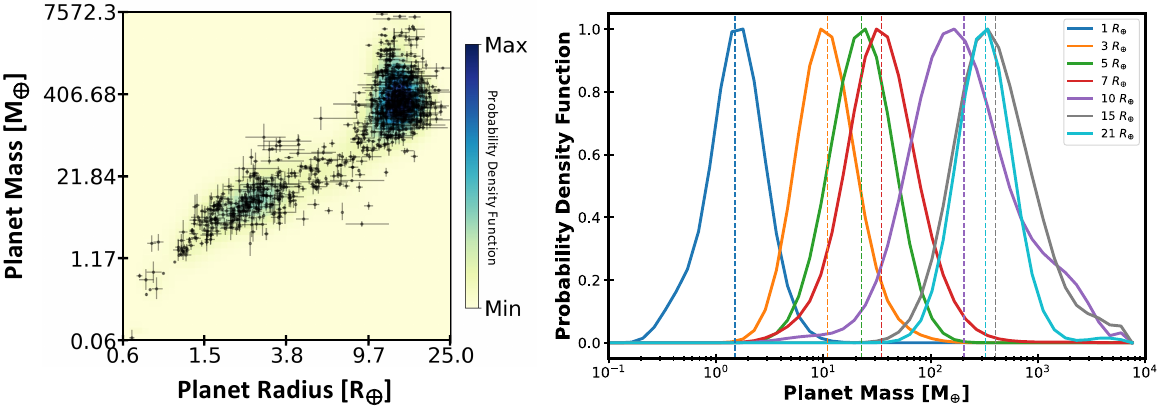}
\caption{Left: Two-dimensional input sample set with the 917 confirmed planets that meet our inclusion criteria of having published mass and radius measurements at the 50\% or better level. Background color shows the corresponding joint probability distribution $f(m, r)$, where blue is a higher probability and yellow is lower. Right: Conditional mass distribution $f(m|r)$ for the joint mass-radius distribution for an example set of planet radii. The vertical dashed lines denote the expected mass value while the curve shows the probability density function for each radius.}
\label{fig:MR_Relation}
\end{figure*}

When selecting transiting exoplanets for atmospheric characterization, there are two relevant populations of potential targets. The first is the set of confirmed, transiting exoplanets whose existence and some subset of physical properties have been verified via follow-up observing efforts and then published in the peer-reviewed literature. To capture the current state of known exoplanets, we queried the NASA Exoplanet Archive's Planetary Systems Composite Data Catalog \citep{10.26133/NEA13}\footnote{Accessed May 21, 2025} and obtained a list of all ``confirmed exoplanets'', which includes statistically validated planets. We remove any known planets whose parameter set does not include at least an orbital period, a transit epoch, and uncertainties for both of these values. We note that this particular table can combine stellar and planet properties from multiple published results and that the values presented for a single object may not be self-consistent. But as we are interested in target lists that span many hundreds objects we accept this lack of self-consistency in favor of obtaining a more complete set of properties for each planet.

The second population is the set of exoplanet candidates that have not yet been confirmed via follow-up observations. To incorporate these objects, we downloaded the current list of TESS Objects of Interest (TOIs) from the TESS Exoplanet Vetter (TEV) website\footnote{https://tev.mit.edu, accessed May 21, 2025}. We supplement this list with the most recent TOI disposition for each candidate from the Exoplanet Follow-up Observing Program (\nobreak{ExoFOP}) website\footnote{https://exofop.ipac.caltech.edu/tess/}. These dispositions indicate whether individual TOIs have been identified as false positives or remain potential planet candidates as additional follow-up observations are taken by the community. As a first step we remove any TOIs that have dispositions indicating that the transit signal is due to a false positive, that the TOI's orbital period and/or transit epoch is unknown\footnote{The full list of dispositions that resulted in removal from the TOI list for is: APC, BD, BEB, EB, FA, KP, LEPC, NEB, NPC, P, PNEB, SB1, SB2, SEB, SEB1, STPC, and APC.}. To the TESS targets, we add the Kepler Objects of Interest (KOIs) included in the Cumulative KOI Table \citep{10.26133/NEA4}\footnote{Accessed June 5, 2024} hosted at the Exoplanet Archive. This table compiles information from the individual KOI activity tables hosted at the Exoplanet Archive that describe the current results of different searches of the Kepler light curves.

We remove any TOIs or KOIs that have radii measured to be larger than two Jupiter radii or equilibrium temperature estimates hotter than 2565~K\footnote{The 99.7$^{\rm{th}}$ percentile of the \Teq\ distribution among known exoplanets on the Exoplanet Archive} as these are likely to be eclipsing binaries or brown dwarfs. We then remove any TOIs or KOIs are already listed as confirmed planets in the list downloaded from the Exoplanet Archive to prevent the double-counting of planet candidates and the known planets lists.

\subsection{Estimating Planet Masses}\label{sec:MR_relation}

We use the open source \texttt{MRExo} package \citep{Kanodia2019, Kanodia2023} to determine a non-parametric exoplanet mass-radius relation and then assign masses to any of the objects in our combined known planets + TOIs + KOIs list that don't already have published mass measurements. This formalism utilizes a convolution of beta and normal distributions to fit a joint mass-radius --- $f(m,r)$ --- probability distribution which can then be conditioned to predict masses from given radii or vice versa \citep{Ning2018}. We assemble an input dataset from which to derive the mass-radius relation from the NASA Exoplanet Archive's Confirmed Planets Table \citep{10.26133/NEA12}\footnote{Accessed December 13, 2023} by selecting the default parameter sets for all planets listed as ``Published Confirmed'' that contain both mass and radius measurements measured to at least 50\% precision. This restriction reduces the scatter among the input data while minimizing the detection bias that can skew small planet mass-radius relations if such analyses only consider high precision (e.g., $\geq$5$\sigma$) mass measurement results \citep[see][]{Burt2018,Montet2018}. The resulting input dataset contains 917 planets that have radii from 0.64 to 22.75 R$_{\oplus}$ and masses from 0.07 to 6884 M$_{\oplus}$, which in turn sets the mass and radius parameter space in which \texttt{MRExo} can predict a planet's mass given its radius. (Figure~\ref{fig:MR_Relation}). We use the cross validation method to optimize for the number of degrees (i.e., the complexity) of the beta distributions \citep{Kanodia2023} and obtain 43 and 47 degrees for the radius and mass axes, respectively. Subsequently we fit for weights for each of these beta distributions using the \lq{}majorize-minimization\rq{} (MM) algorithm and then use this joint distribution to assign a mass to each planet in our list of viable targets that does not have a published mass measurement.

\subsection{Estimating Orbital Eccentricity \& Longitude of Periastron}\label{sec:EccOmega}

For planets that do not have orbital eccentricities and longitudes of periastron reported as part of their composite parameter set on the Exoplanet Archive we follow the prescription of \citet{Shabram2016} to determine a simulated $e\, \rm{cos}(\omega)$ for that planet. \citet{Shabram2016} determined that the $e\, \rm{cos}(\omega)$ distribution of transiting, short period (P \textless 50 days) exoplanets is best described by a two-component Gaussian mixture model where $\sim$89\% of planets belong to a low-eccentricity population with $\mu = 0$ and $\sigma$ = 0.01 while the remaining $\sim$11\% of planets belong to a higher eccentricity population with $\mu = 0$ and $\sigma$ = 0.22. For each target that does not have published eccentricity and $\omega$ values, we draw a random number between 0 and 1 and assign those with N $\leq$ 0.89 to the low eccentricity group and those with N \textgreater 0.89 to the high eccentricity population. We then draw an $e\, \rm{cos}(\omega)$ value from the corresponding distribution and assign each object an $\omega$ value drawn randomly from a uniform distribution from 0-359 degrees. Using these two values, we back out the corresponding orbital eccentricity value and add all parameters to the target list.

\subsection{Assigning Mass Precision Requirements \& Corresponding Cuts}\label{sec:Mass_Needs_Cuts}

The accuracy to which the atmospheres of transiting exoplanets can be understood relies on precise knowledge of their surface gravities, which comes from precise measurements of their radii and masses. Transiting planets generally have precise radius measurements (the median uncertainty in planet radius among the default parameter sets of confirmed transiting planets on the Exoplanet Archive is 9.8\%), which means that the uncertainty in their surface gravities is most often dominated by the uncertainty in their mass measurements. \citet{Batalha2019} demonstrates that for small (R$_{\rm{p}}$ \textless 4 R$_{\oplus}$), temperate planets a mass precision of 5$\sigma$ or better is necessary to ensure that the uncertainty in the atmospheric retrievals is not dominated by the uncertainty in the planet's mass. For larger planets, and planets that are both smaller than Neptune and hotter than T$_{\rm{eq}} = 900$ K, the limiting mass precision is instead 2$\sigma$.

We assign each planet a mass uncertainty requirement based upon its size and equilibrium temperature, either 2$\sigma$ or 5$\sigma$ following the \citet{Batalha2019} guidelines. For those planets that are identified as members of our eclipse spectroscopy target list we increase the mass precision requirement to 10$\sigma$ in an effort to constrain the planet's orbital eccentricity (see Section \ref{section:rep_emiss} for further discussion on this) which is key to determining future eclipse ephemerides. We note that while our mass uncertainty requirements range from 2 to 10$\sigma$, the mass precision required to answer a given scientific hypothesis depends on both the question being asked and the sample size of planets being considered. The planet catalog used to derive a mass-metallicity relation among giant planets in \citet{Thorngren2016}, for example, contains 42 objects ranging from 20--700 M$_{\oplus}$ and has a median mass uncertainty of 12$\sigma$.

\begin{figure}[ht!]
    \centering
    \includegraphics[width=.45\textwidth]{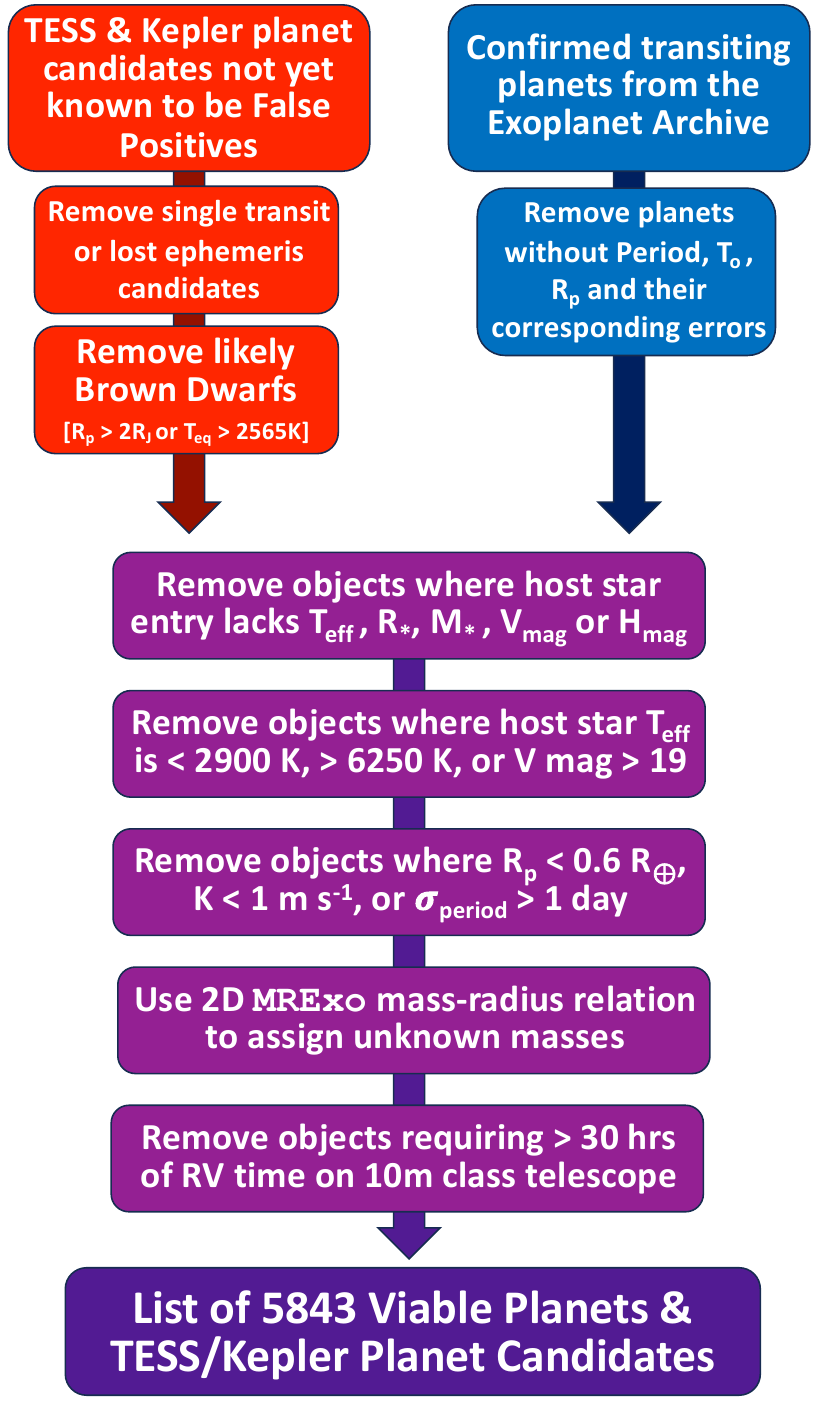}
    \caption{Visual summary of the target list assembly process, starting from a list of confirmed transiting planets downloaded from the Exoplanet Archive and the lists of TESS and Kepler Objects of Interest (TOIs/KOIs) that have not yet been found to be false positives via ground- and space-based follow up observing efforts.}
    \label{fig:TargetListCreation}
\end{figure}

Exoplanet masses are generally measured via radial velocity (RV) observations of the host star, which are taken with high resolution, ground-based spectrographs \citep[see, e.g.,][for a review]{Fischer2016}. Precise RV measurements generally require the planet's host star to be below the Kraft break \citep[\Teff\ $\leq$ 6250 K][]{Kraft1967,Beyer2024} to ensure there are sufficient absorption lines in the stellar spectra and that those lines are narrow enough (not subject to significant rotational broadening) to provide the RV information content necessary to measure precise gravitational reflex motions within the spectrum \citep{Beatty2015}. As we highlight the need for masses to support atmospheric spectral retrievals in this paper, we remove any planets and planet candidates that currently lack sufficient mass measurements and orbit stars with \Teff\ $\geq$ 6250 K from the target list. 

Finally, we implement a broad cut on the amount of RV resource time necessary to obtain the recommended mass precision. The details of these calculations are provided in Section \ref{sec:RVTime} and are built around the analytic expressions from \citet{Cloutier2018}. We remove any planets or planet candidates whose host stars lack measurements of their effective temperature, mass, radius, or V-magnitude, as these values are necessary to calculate the RV time requirements. We also remove objects whose host stars are cooler than 2900 K, i.e. below the main sequence, or have a V-magnitude \textgreater\ 19, as these lay outside our RV simulation capabilities. We then compute RV semi-amplitudes for all objects that remain on the viable targets list and remove any planets or planet candidates whose expected RV semi-amplitudes are below 1 m s$^{-1}$ as modern RV spectrographs can still struggle to reliably and robustly detect Keplerian signals at this level, especially in the presence of stellar variability \citep[see, e.g.,][]{Zhao2022}. Finally, we remove any planets or planet candidates that would require more than three full nights of time ($>$30~hours) on a 10-meter class telescope to acquire their recommended 2-, 5-, or 10$\sigma$ mass precision.

\subsection{The Full List of Viable Targets}\label{sec:Full_Target_List}

\begin{figure*}[htb!]
    \centering
    \includegraphics[width=.90\textwidth]{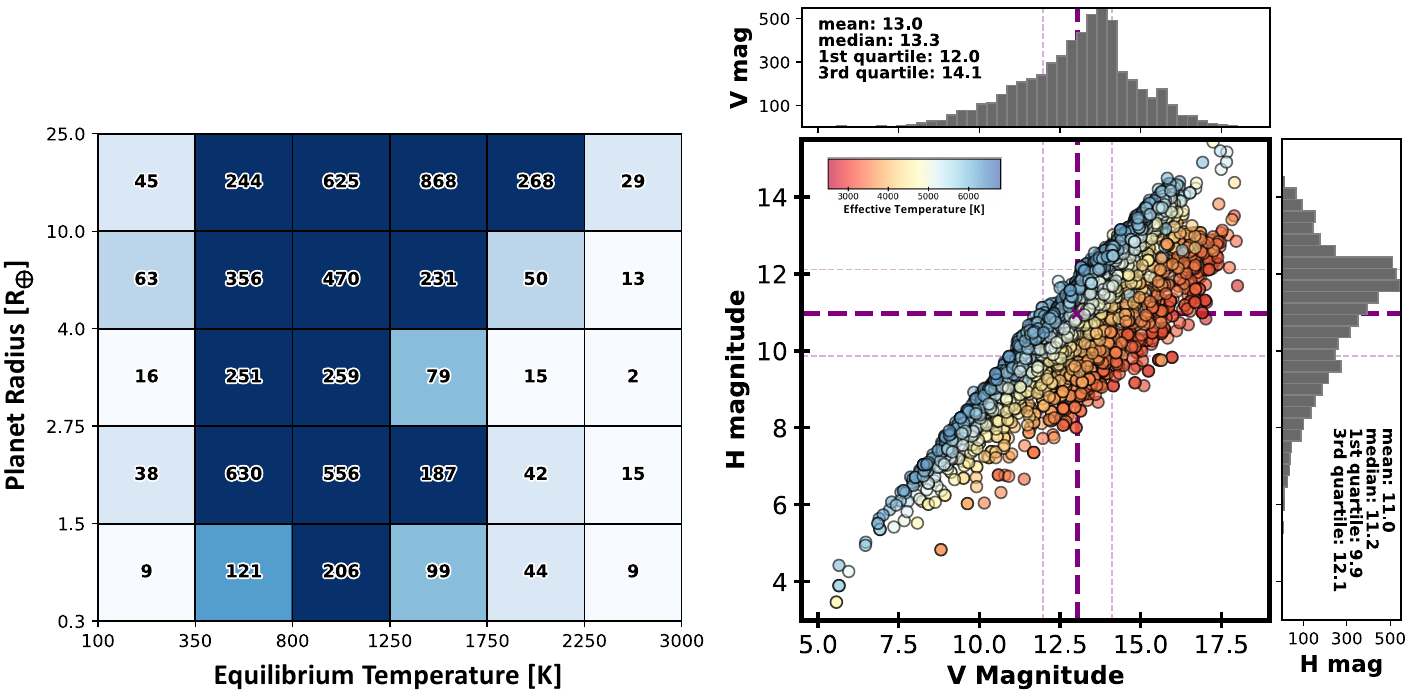}
    \caption{Left: The 5843 known planets and TESS/Kepler planet candidates that meet all of the selection criteria (summarized in Figure \ref{fig:TargetListCreation}) spread across the equilibrium temperature and planet radii bins standardized in \citet{hord23}. Larger planets dominate the population, but thousands of sub-Neptune sized planets are present in this initial list. Right: Host stars for these same known planets and TESS/Kepler planet candidates. Bold purple dashed lines denote the mean V and H band magnitudes, while the thinner purple dashed lines denote the 1$^{\textrm{st}}$ and 3$^{\textrm{rd}}$ quartiles in each magnitude distribution.}
    \label{fig:Full_Stellar_Sample}
\end{figure*}

This selection process is captured in Figure~\ref{fig:TargetListCreation} and the resulting population of viable planets and planet candidates and their host stars is depicted in Figure \ref{fig:Full_Stellar_Sample} where we visualize the planet population by spreading its members across the planetary radius and equilibrium temperature grid established in \citealt{hord23}. In total we end up with 5843 viable targets, 2276 (39\%) of which are known planets while 3567 (61\%) are planet candidates from either TESS (54\%) or Kepler (7\%). While the ratio of input planet candidates from TESS and Kepler is just under 4:1, the number of candidates that makes it all the way through to the list of viable targets is just over 8:1 in favor of the TESS candidates (89\% TOIs vs 11\% KOIs). This 2x increase is driven primarily by the requirement that the candidate planet can obtain its assigned mass precision in fewer than 3 nights on a 10-m class telescope, which rules out 73\%\ of KOIs but only 2\% of TOIs that make it through all of the earlier selection criteria.

The host stars range in brightness from 3.5 to 15.5 magnitude in H-band and 5.5 to 18.0 magnitude in V-band. All allowed spectral types (those with \Teff\ from 2900--6250~K) are represented, with the list being dominated by G and K stars which make up 40\% and 27\% of all of the viable host stars, respectively. F stars account for another 23\% of the list, while M stars comprise only 9\% of the potential target hosts. The M dwarf percentage is, however, slightly higher than the corresponding percentages of M dwarfs from the three input lists (6.7\% M dwarfs) or after applying all of the selection criteria other than the 3 night RV time cut (8.0\% M dwarfs). The relative increase in M dwarf hosts is driven by the fact that the TESS planet candidates make up a majority of the viable targets list as they orbit brighter stars and have larger semi-amplitudes, making them more likely to survive the 3 night RV cut, and TESS is optimized to detect planets around M dwarf stars.

\section{Generating an Exoatmospheric Target List}\label{sec:RepTargetList}

While the total number of confirmed exoplanets is now over 5,900 there remain many regions of the two dimensional \Teq\ and planet radius parameter space that lack a significant number of planets (or even planet candidates) to study (Figure \ref{fig:Full_Stellar_Sample}). One such dearth of planets occurs at very high \Teq\ values, likely due to the dynamical interactions required to migrate planets into ultra short period orbits \citep[see, e.g.,][]{Millholland2020}. Another dearth, more likely driven by observational biases rather than a true population limitation, occurs at smaller planet radii and longer orbital periods (corresponding to lower \Teq\ values) as these planets have smaller transit depths and lower transit probabilities, respectively \citep{winn10}. For the smaller and longer period planets that do transit, follow up confirmation efforts to observe additional transits and/or determine the planets' masses are also more challenging and generally require more resources than larger and/or shorter period planets. 

Despite these challenges, characterizing a variety of planets per grid space is crucial as numerous physical parameters beyond planet size and equilibrium temperature can influence a planet's atmospheric spectrum. These include the planet's mass, atmospheric composition, metallicity, and internal heat budget, to name just a few. Indeed, results from large HST surveys show how even hot Jupiters that share similar sizes and equilibrium temperatures can have large variations in chemical abundances and cloud/haze properties \citep{sing2016}. Obtaining a sense of what atmospheric characteristics are common within a given radius and temperature bin thus requires having a large enough population of well characterized planets to be able to identify both commonalities and outlier behaviors.

\citealt{kempton18} notes that simulations performed by the FINESSE Science Team \citep{FINESSE2017} found that on order 500 planets are required to identify statistical trends between stellar and planet parameters when accounting for the diversity of planet outcomes predicted by formation models \citep{fortney13,Mordasini2016}. And the Ariel science team intends to target roughly 1,000 planets in transmission spectroscopy to support the mission's comparative planetology goals \citep{edwards2022}. For this work we select a middle ground of 750 transmission spectroscopy planets as, in an ideal scenario, this would allow for a sufficient number of confirmed and well characterized planets to be spread evenly across the 30 grid spaces in our \Teq\ versus \Rearth\ parameter space resulting in 25 planets per bin. 25 planets per bin enables straightforward $\sqrt{N}$ statistics and, depending on the planets, can allow for comparisons along additional axes such as metallicity or stellar host type. We then set the total number of eclipse targets to be 150, five times smaller than our transmission spectroscopy target list. This in keeping with current and planned future exoplanet atmospheric characterization efforts. A recent analysis of time allocated through the JWST Guest Observer program shows that the ratio of hours awarded for transmission to eclipse spectroscopy ranges from $\sim$15-25\% over Cycles 1-4 \citep{Espinoza2025}. And current Ariel survey plans also adopt a 20\% ratio, aiming to observe 200 exoplanets in eclipse \citep{edwards2022}.

In neither the transmission nor emission case, however, does the current census of confirmed planets allow for such an ideal spread across the \Teq\ and \Rearth\ axes. And so we must consider alternative approaches.

\subsection{Selecting Targets Via a Figure of Merit}\label{section:fom_trans}

A traditional approach to survey design is to rank objects of interest using a Figure of Merit (FOM) and then cull the highest scored objects to form the survey's target list. To capture this methodology we calculate a telescope-agnostic transmission spectroscopy figure of merit (FOM$_{\mathrm{transit}}$, as defined in \citealt{zellem17} and independently derived by \citealt{cowan15}, \citealt{goyal18}, \citealt{kempton18}, and \citealt{morgan19}) for each planet and planet candidate in our list of viable targets. Adopting the convention in \citet{zellem17}, this metric scores targets based upon their predicted spectral modulation and host star brightness:
\begin{align} \label{eqn:FOM_transit}
    FOM_{\mathrm{transit}} = \frac{2H_{s}R_{p}R_{s}^{-2}}{10^{0.2H_{mag}}}
\end{align}
where $R_{p}$ and $R_{s}$ are the planet and host star radii, respectively, and $H_{mag}$ is the host star’s H-band magnitude. We select the H-band because it is centered on 1.65$\micron$ which is accessible to JWST, aligns with Ariel's peak photon conversion efficiency \citep{mugnai20}, and corresponds to the peak black body emission of 1800 K object, near the center of our \Teq\ temperature range. 

To estimate the planet's scale height, $H_{s}=\frac{kT_{eq}}{mg}$, we calculate its acceleration due to gravity $g$ from its observed radius and its measured or predicted mass. We then either adopt the \Teq\ value reported by the NASA Exoplanet Archive or, in cases where the Archive lacks an entry for \Teq, calculate it as \Teq~=~\Teff\ $\sqrt{R_{*}/a} (\frac{1}{4})^\frac{1}{4}$ which assumes zero albedo and full day-night heat redistribution. We estimate the planet's mean molecular weight and assign each planet a metallicity based on an assumed mass-metallicity relationship while adopting a solar C/O value of 0.58 (log$_{10}$(C/O) = $-$0.26), following the prescription detailed in \citet{zellem17}.

We construct two target lists, one that draws only from the planets that have been confirmed in the published literature (\lq{}Known Planets Only\rq{} in all following figures) and one that draws from both the confirmed planets and the TESS and Kepler planet candidates (\lq{}Known Planets + Candidates\rq{}). In each case we select only the 750 top ranked planets / planet candidates. Unsurprisingly, this approach results in target lists that are largely biased towards hot Jupiters and super-Jupiters (left column of Figure \ref{fig:popstats_transmission}). These larger, hotter planets have significantly larger spectral modulations thanks to their extended atmospheres, low mean molecular weights, and low surface gravities, which makes them excellent candidates for transmission spectroscopy efforts. Comparing these populations to the full list of viable candidates (Figure \ref{fig:Full_Stellar_Sample}) it becomes clear that the cost of prioritizing these more easily accessible transmission targets is a corresponding neglect of the smaller and/or cooler planet regions of the \Teq\ vs. \Rearth\ parameter space.

\subsection{Generating a More Representative Target List}\label{sec:rep_trans}

\begin{figure*}[ht!]
    \centering
    \includegraphics[width=.95\textwidth]{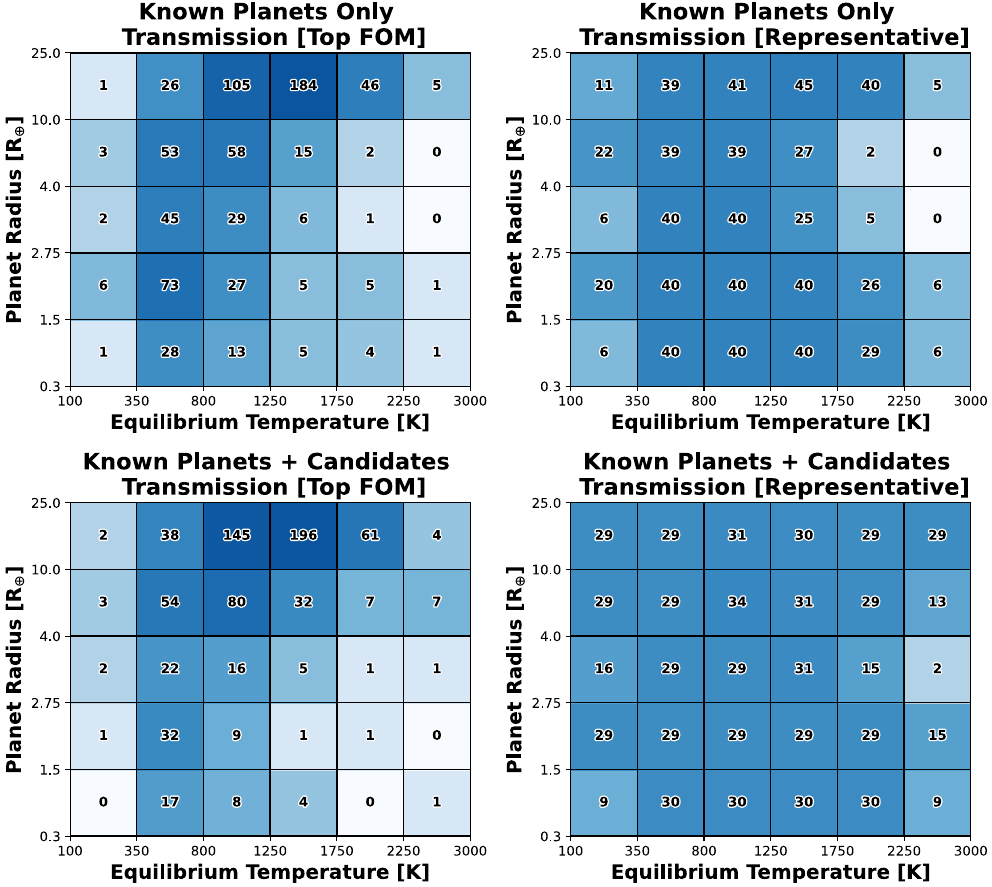}
    \caption{Planetary radii and equilibrium temperatures of the four transmission spectroscopy planet populations described in Sections \ref{section:fom_trans} \& \ref{sec:rep_trans}. The left column takes the top 750 objects from the list of viable targets as ranked by a platform-independent transmission FOM (Eqn.~\ref{eqn:FOM_transit}), while the right column populations are assembled by iteratively passing through the grid and taking the top ranked object in each radius/\Teq\ bin until the target list reaches 750 objects. The top row populations select targets from the confirmed planets alone, while the bottom row populations draw from both the confirmed planets and the TESS and Kepler planet candidates.}
    \label{fig:popstats_transmission}
\end{figure*}

To establish a more representative target list that evenly samples the variety of available planet types, we make active use of the \citet{hord23} \Rearth\ vs \Teq~grid, this time developing the target list such that the grid spaces are filled as evenly as possible for as long as possible on the way to reaching 750 targets. Beginning in the bottom left grid corner (which contains planets from 0.3--1.5~R$_{\oplus}$ and with \Teq~from 100--350~K) we select the planet with the highest transit figure of merit ranking $FOM_{\mathrm{transit}}$ \citep[Eqn.~\ref{eqn:FOM_transit};][]{zellem17} in that bin and add it to the target list. We then move to the next bin and select the highest $FOM_{\mathrm{transit}}$ in that region of \Rearth\ and \Teq\ space and add it to our target list. This process is repeated until we have run through every single bin, at which point we return to the first bin and select the next highest ranked target within that bin to include on the target list. If all targets within a given bin have already been added to the target list in earlier passes through the grid, then that bin is skipped over. This process continues until the total number of targets reaches 750. 

We again construct two target lists following this methodology, one that draws only from the planets that have been confirmed in the published literature (\lq{}Known Planets Only\rq{}) and one that draws from both the confirmed planets and the TESS and Kepler planet candidates (\lq{}Known Planets + Candidates\rq{}). The right column of Figure~\ref{fig:popstats_transmission} illustrates the result of this target list construction approach: a set of transmission spectroscopy targets where the 750 planets are distributed much more evenly over the \Teq\ vs \Rearth\ parameter space than when target selection depends purely on the targets' ranked $FOM_{\mathrm{transit}}$ values. 

\subsection{Generating a Representative Emission Spectroscopy Target List}\label{section:rep_emiss}

For emission, or similarly phase curve, observations \citep[e.g.,][]{harrington06, cowan07, cowan12, knutson07, Knutson_2012, Lewis_2013} we apply a figure of merit FOM$_{\mathrm{eclipse}}$, first defined in \citet{zellem18} and later independently derived in \citet{kempton18}, that provides a relative ranking of targets based upon their predicted eclipse depth and host star brightness:
\begin{align} \label{eqn:FOM_eclipse}
    FOM_{\mathrm{eclipse}} = \frac{F_{p}R_{p}^{2}F_{s}^{-1}R_{s}^{-2}}{10^{0.2H_{mag}}}
\end{align}
\noindent where $F_{p}$ and $F_{s}$ are the fluxes of the planet and host star, respectively, estimated by the Planck function calculated at the center of the H-band (1.65~$\mu$m). 

Many of the \Teq\ and \Rearth\ regimes sampled in our population grid would not be observable in emission from any modern spectroscopy platform, specifically the smaller and/or cooler regions of the grid. For example, a temperate Earth-like planet (\Teq\ = 288 K) around an M-dwarf star (\Teff\ = 2000 K; R$_{p}$/R$_{s}$ = 0.1) will have an eclipse depth of only 5$\times 10^{-8}$~ppm at 1.65~$\mu$m whereas Ariel's currently-adopted noise floor is significantly larger at 20~ppm \citep{mugnai20}. Recognizing this limitation, we again start from the full list of 5843 viable planets and planet candidates, but this time remove any whose eclipse depth at 1.65$\mu$m is predicted to be less than 20~ppm. We also set the mass requirement for these potential eclipse targets to 10$\sigma$ as for planets with small orbital eccentricities the RV signal attributable to the orbital eccentricity is well approximated by the epicyclic approximation, i.e., a sinusoidal signal with amplitude $e$K \citep{Luhn2023}. A mass precision of 10\% then also implies an eccentricity precision of 0.1 and, as discussed in Section \ref{sec:eclipse_times}, knowledge of a planet's eccentricity plays an important role in decreasing the uncertainty in future eclipse ephemerides. We follow the RV time estimate procedure detailed Section \ref{sec:RVTime} and discard any planets where reaching a 10$\sigma$ mass uncertainty would require more than 3 full nights of time on a 10-m class telescope.

\begin{figure}[ht!]
    \centering
    \includegraphics[width=.4\textwidth]{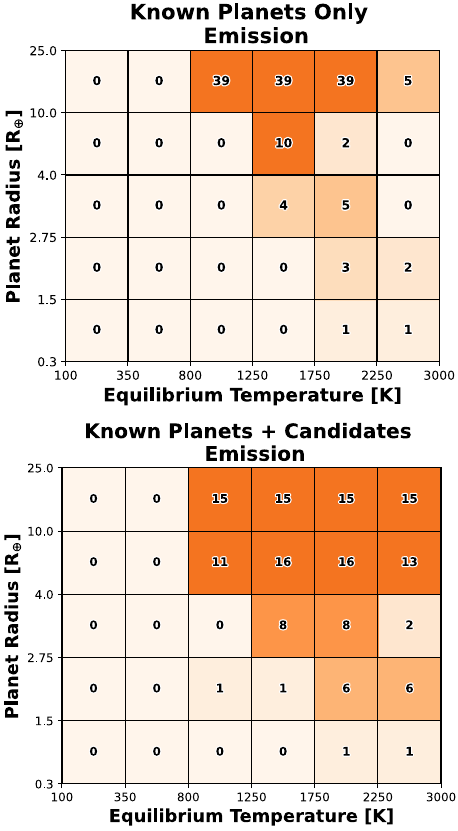}
    \caption{Planetary radii and equilibrium temperatures of our 150 planet representative emission spectroscopy populations, assembled following our representative target list methodology. The top plot select targets from the known planets alone, while the bottom plot draws from both the known planets and the TESS/Kepler planet candidates. Only planets / planet candidates with an estimated eclipse depth $\geq$ 20 ppm in H-band (1.65 $\mu$) are considered and which can obtain a 10$sigma$ mass uncertainty in less than 3 nights of Keck time are considered. Drawing from the planet candidates produces numerous grid spaces with more than 5 objects present leading to a more representative survey than what would be possible when only considering known planets.}
    \label{fig:popstats_emission}
\end{figure}

Once our list of viable eclipse planets and planet candidates has been assembled, we repeat the representative target list methodology and again cycle through the \Teq\ and \Rearth\ grid adding the top ranked FOM$_{\mathrm{eclipse}}$ object from each grid space and skipping the grid space if no viable targets remain. Figure \ref{fig:popstats_emission} shows the resulting emission planet populations when drawing only from the list of previously confirmed planets and when drawing from both confirmed planets and planet candidates. As expected, there is a dearth of small and/or cool planets present in both lists, as their eclipse depths are too small for modern facilities to detect. Considering both known planets and planet candidates, however, does help to redistribute the targets down from being concentrated in the super-Jupiter regime that is favored by the known planets only list.

\subsection{Moving Forward with the Representative Target Lists}\label{section:rep_lists}

\begin{figure*}[htb!]
\includegraphics[width=.95\textwidth]{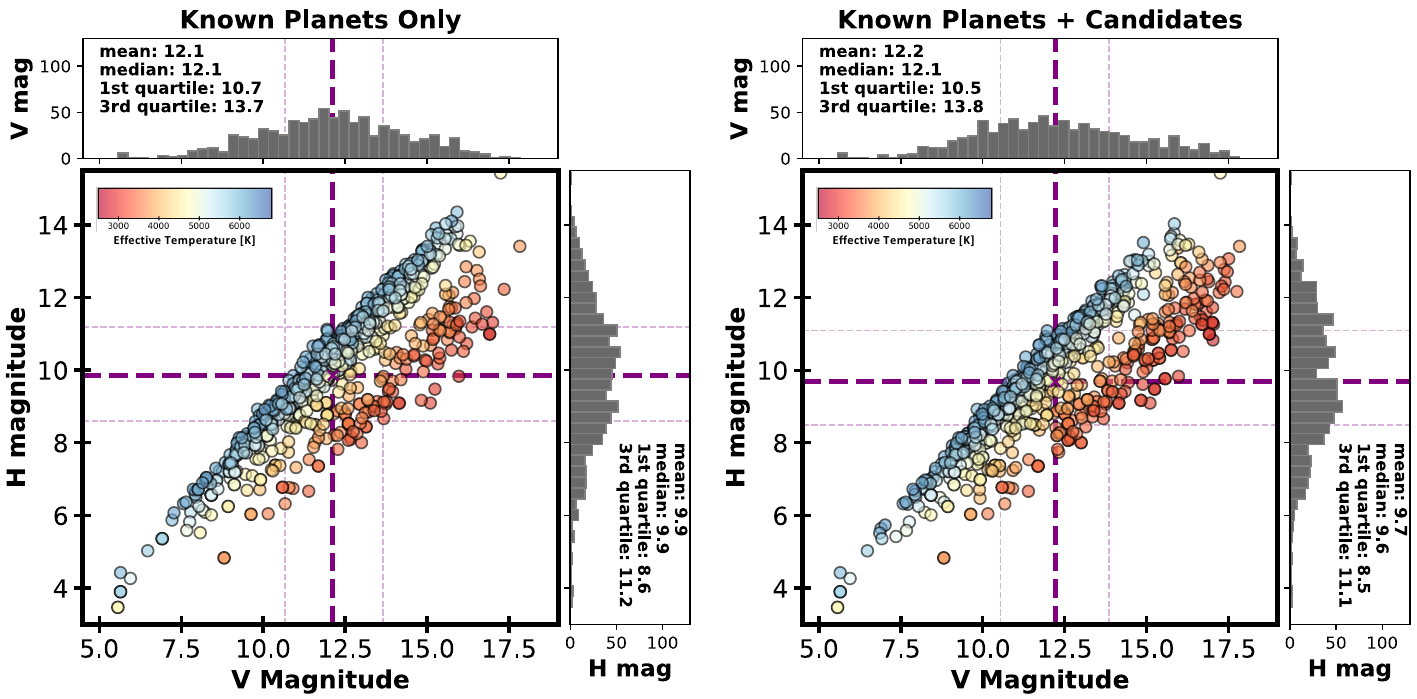}
\caption{Planet host star distributions for each of the representative transmission + emission spectroscopy target lists described above, following the same layout as in Figure \ref{fig:Full_Stellar_Sample}. Adding in the TESS and Kepler candidate planets alongside the known planets (right plot) shifts the stellar host population to include more M stars than the known planets only lists, unsurprising given that the TESS mission was optimized to detect planets around cool, nearby stars.}
\label{fig:StellarHists}
\end{figure*}

We combine the the 750 object transmission spectroscopy list and the 150 object emission spectroscopy list into a single, primary list that contains both the transmission and emission targets, removing duplicates of planets that were selected for both observation types and keeping the more stringent mass measurement requirement in these cases. The resulting target lists contain 759 objects in the known planets only case, and 762 objects when we consider both known planets + planet candidates. In both cases the majority of the 150 emission targets are also selected for transmission spectroscopy with only 9 and 12 objects selected for emission spectroscopy alone in the known planets and planets + candidates cases, respectively. Both target lists sample planets around all allowed host star types (those with \Teff\ from 2900--6250~K) as evident in Figure \ref{fig:StellarHists}. Given the increased diversity of planet types, we adopt these representative lists as our preferred survey approach and proceed to quantify the amount of ground-based photometric and RV characterization work needed to prepare the planets and planet candidates for future atmospheric spectroscopy efforts.

\section{Quantifying the Uncertainty in Predicted Transit and Eclipse Timings}\label{sec:TransitUncert}

To ensure observing efficiency, the mid-transit and mid-eclipse times of all of the targets on our representative target lists should be known to high precision. Yet many of these targets have \lq{}stale\rq{} transit ephemerides, where the 1$\sigma$ uncertainty in the mid-transit time $\Delta T_{mid}$ exceeds half the transit duration, which can result in the loss of significant observing time \citep{zellem20}.

\begin{figure}[htb!]
\includegraphics[width=.48\textwidth]{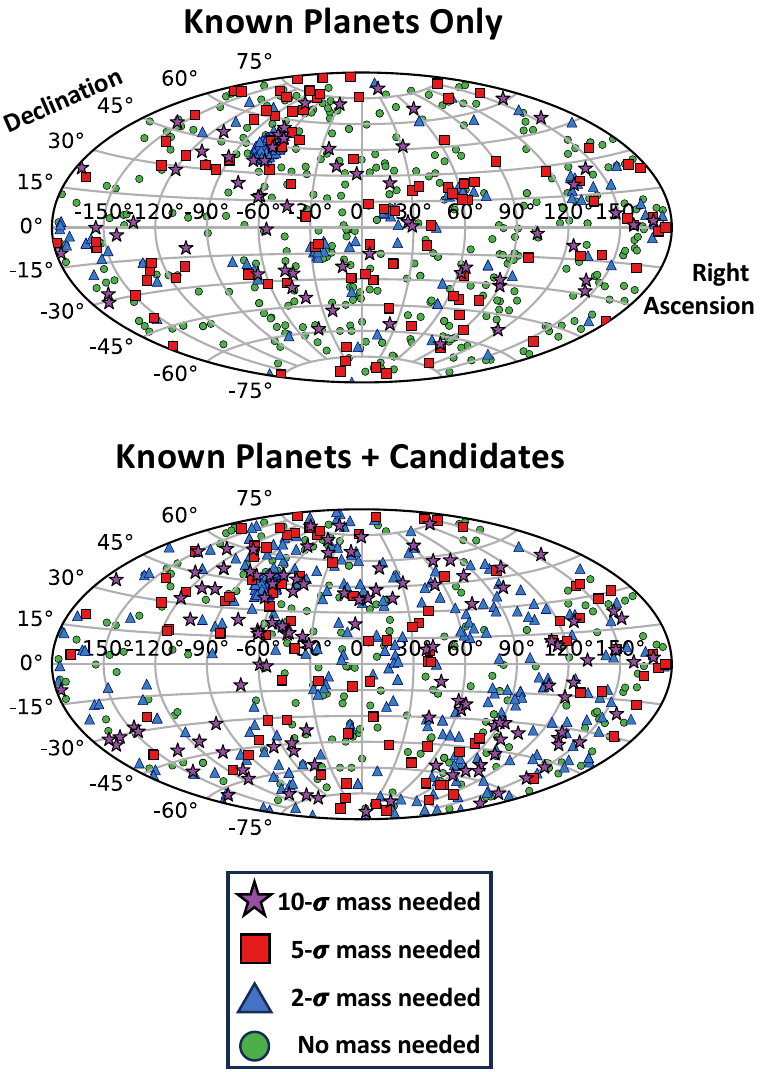}
\caption{Distribution of the representative target lists when considering only those planets that have been previously confirmed (top plot) and confirmed planets + current TESS/Kepler planet candidates that have not yet been found to be false positives (bottom plot). Color coding indicates whether the planets require a 10$\sigma$ (purple stars, reserved for eclipse targets), 5$\sigma$ (red squares, for small and temperate transmission targets), or 2$\sigma$ (blue triangles for large and/or hot transmission targets) mass measurement, or if they already have a sufficient mass measurement (green circles).}
\label{fig:SkyMap}
\end{figure}

Any precursor observing effort focused on refining the transit times and/or masses of these targets will require a global effort, as the objects on our target lists cover a broad range of right ascension and declination values (Figure~\ref{fig:SkyMap}). This need is especially true when the current TESS planet candidates are considered as this results in a more homogeneous spread of targets across the sky, rather than the tight clustering within the Kepler and K2 fields that is evident when only considering confirmed planets. 

\subsection{Mid-Transit Timing Uncertainties}\label{subsec:transit_times}

We compute the uncertainty in a planet's mid-transit time $T_{mid}$ via the following equation, as analytically derived in detail in \citet{zellem20}:

\begin{align}
\label{eqn:Tmid_err}
    \Delta T_{mid} = &(n_{orbit}^{2} \cdot \Delta P^{2} \\
    &+ 2n_{orbit} \cdot \Delta P \Delta T_{0} \nonumber \\
    &+ \Delta T_{0}^{2})^{1/2} \nonumber
\end{align}

where $n_{orbit}$ is the number of planetary orbits that have elapsed since an ephemeris reference time $T_{0}$ and $P$ is the exoplanet's orbital period. Following the procedure detailed in \citet{zellem20} we calculate the uncertainty of each planet's mid-transit time for current atmospheric characterization missions such as HST and JWST (adopting a reference date of 1 July 2025, corresponding to the start of JWST Cycle 4) and also future missions (adopting 1 May 2030 as a potential start of Ariel science operations). 

\begin{figure}[htb!]
\includegraphics[width=.48\textwidth]{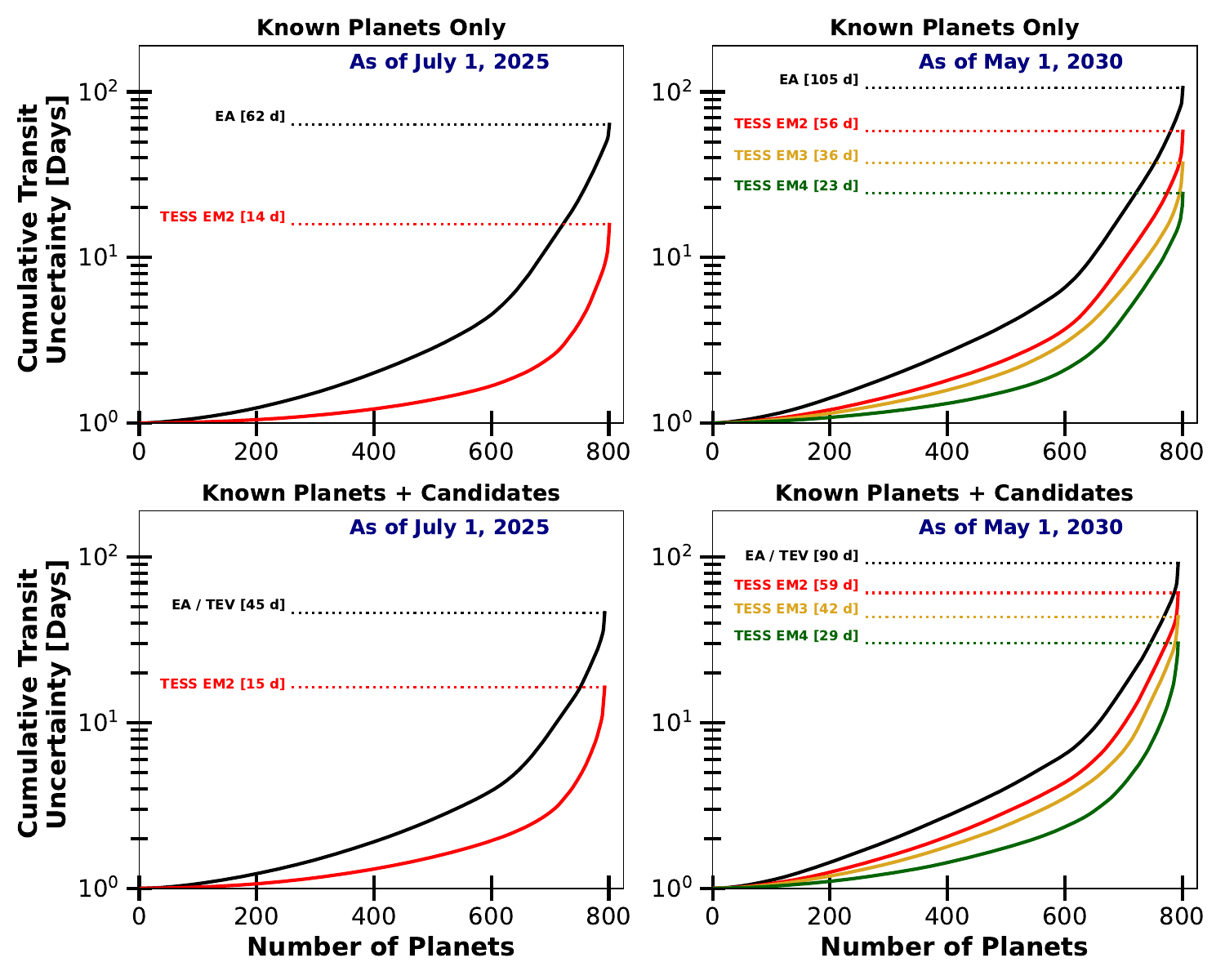}
\caption{The cumulative uncertainty for the mid-transit times of the two representative target lists at the JWST Cycle 4 start date of July 1, 2025 (left column) and a predicted Ariel start date of May 1, 2030 (right column). In the worst-case scenario, where no transit ephemerides are refreshed beyond what is currently reported in the Exoplanet Archive (black lines) there will be over 40 days of uncertainty for these targets during JWST's Cycle 4 operations and over 90 of days of uncertainty by the time Ariel begins science operations. Observations by TESS's current extended mission (EM2, red lines) and our suggested extended missions (EM3 in gold, and EM4 in green) all decrease the total uncertainty, and by EM4 are able to bring it down to 23-29 days for Ariel.}
\label{fig:Tmid_Uncertainty}
\end{figure}

For the known planets we use the orbital period, transit ephemeris, and associated uncertainties from each object's composite parameter set on the NASA Exoplanet Archive. For the TESS and Kepler planet candidates we use the orbital period, transit ephemeris, and associated uncertainties presented in the TESS Exoplanet Vetter and Cumulative KOI tables. We note that these orbital parameters are not always the most recent, or precise, values, and so the results presented here are conservative. In the worst-case scenario, where none of the orbital parameters are refreshed beyond what's presented in the Archive, TEV, and cumulative KOI catalogs, we find, using Equation~\ref{eqn:Tmid_err} and following the prescription in \citet{zellem20}, that 45--62 days of JWST time could be wasted if the survey began in Summer 2025, and 90--105 days of Ariel's observing time could be wasted during a survey carried out in 2030 (Figure~\ref{fig:Tmid_Uncertainty}).

A more realistic view of the transit ephemeris refreshment effort includes future contributions made by the TESS mission, which continues to survey the sky and provide light curves of most nearby stars. TESS' second extended mission (EM2) is currently underway and its observing pattern has been defined through September 2025. We use the \texttt{TESS-point} python package \citep{tesspoint} to predict what EM2 sectors will contain each planet or planet candidate on our target list, and then determine the likely final transit epoch observed for each planet by assuming a 0.9 duty cycle for TESS (TESS team, priv. comm.). That is, we determine the final transit date for each target based upon its period, transit epoch, and list of TESS EM2 sectors during which it will be observed. We assert that this specific transit has a 90\% probability of being captured by TESS and draw a random number from 1-100 to determine the outcome. If the transit was not observed, we shift back one orbital period and again give that transit a 90\% chance of detection. This process is repeated until a transit is marked as \lq{}observed\rq{}, at which point we set the corresponding transit epoch as the new \lq{}fresh ephemeris\rq{} date. 

While we expect that the addition of additional TESS transit observations will improve the precision on the period and ephemeris measurements \citep[see, e.g.,][]{dragomir20} we do not alter the stated precision of either value from our target input catalog. Reducing these uncertainties in response to new TESS data in a robust and reliable way would require simulating the actual TESS light curves and carrying out multi-sector transit fits for each of the 750+ planets in our target lists, which is beyond the scope of this work. Rather we take a conservative approach where we assume that TESS \lq{}does no harm\rq{} and simply move the planet's transit ephemeris into the future while maintaining the previous mid-transit time uncertainty, orbital period, and orbital period uncertainty. Our results therefore capture the lower limit of how much future TESS extended missions can improve transit ephemerides and, by extension, of how much time TESS can save for future JWST and Ariel observation planning efforts.

To consider the impact of future TESS extended missions, which have yet to be fully designed and approved but which seem likely given the mission's success to date, we design two additional Extended Missions (EM3 and EM4) that begin immediately after the conclusion of EM2. We adopt the TESS team's suggested Cycle 8 plan to focus on the south ecliptic pole for Sectors 97-107 as the starting point of EM3 \footnote{see https://tess.mit.edu/tess-year-8-observations/} and then append a duplication of both the TESS prime mission's northern hemisphere survey and the EM1 ecliptic plane survey. For simplicity's sake, we set EM4 to replicate the entirety of the TESS prime mission along with the EM1 ecliptic plane survey. Referring back to already executed or planned sectors, EM3 then includes the same pointings and durations as S97 - S107, S14-26, and S42-46 and runs from September 2025 to January 2028. And EM4 mimics the pointings and durations of S1-26 and S42-46, and runs from January 2028 to the end of April 2030. We replicate the transit ephemeris refresh calculations described above, and determine a final transit date for each target in both EM3 and EM4.

When considering the target list compiled using only confirmed planets, incorporating the transit refresh dates from TESS EM2 decreases the expected observing time losses by a factor of 4x when considering Summer 2025 observations with JWST and HST, from 62 days to 14 days. The effect is a less significant when considering the list made up of known planets + planet candidates (decreasing from 45 to 15 days), in part because their transit epoch dates are skewed closer to the current date thanks to the influence of the TESS planet candidates (Figure ~\ref{fig:Tmid_Uncertainty}). 

When considering a 2030 transmission spectroscopy survey EM2 alone decreases the observing losses by a factor of 1.5 and 2x for the known planets + candidates and known planets only cases, respectively. Adding in ephemeris updates from our speculative EM3 and EM4 drives the time lost in 2030 down even further, decreasing it to from 105 days to 56, 36, and 23 days for the known planets only case and from 90 days to 59, 42, and 29 days for the known planets + candidates case as extended missions 2, 3, and 4 are considered (Figure ~\ref{fig:Tmid_Uncertainty}).

While TESS' all sky survey strategy makes it a valuable source of fresh transit ephemerides for thousands of planets and planet candidates \citep{dragomir20}, that same observing approach results in long stretches between visits to much of the sky. Targets not in the ecliptic poles can go two or more years between observations, allowing many orbital periods to elapse between visits and limiting the power of TESS' ephemeris refinements for individual objects. Additional targeted efforts are therefore required to further reduce these observing time losses, which we discuss in detail in Section~\ref{sec:ephemeris_refinement_recommendations}.

\subsection{Eclipse Timing Uncertainties}\label{sec:eclipse_times}

We next estimate the uncertainties in the mid-eclipse times for our emission spectroscopy samples. We note that the NASA Exoplanet Archive does not currently list mid-eclipse times and so we instead have to estimate them from the mid-transit time. Adapting the formula provided in Equation 33 of \citet{winn10} one can estimate the time between a planet's transit and eclipse midpoints via the following linear approximation\footnote{N.B.: This linear approximation can underestimate the mid-eclipse time for eccentric ($e \geq 0.2$) planets. For example, if one were to use it for HD~80606 b \citep[$e$ = 0.93,][]{pearson22}, the mid-eclipse time would be underestimated by $\sim$15~days. However, only 6-7 of the planets on our two target lists have eccentricities above 0.2 and all but one of those eccentricities are $e \leq\ $0.4, so we do not expect use of this approximation to significantly skew our results.}:
\begin{align}
     \Delta t_{c} \approx \frac{P}{2} \bigg[1 + \frac{4}{\pi}e \cos (\omega) \bigg]
\end{align}

\noindent One can then predict a mid-eclipse time from the mid-transit time via:
\begin{align}
    T_{\mathrm{mid\, eclipse}} \approx T_{\mathrm{mid}} + \frac{P}{2} \bigg[1 + \frac{4}{\pi}e \cos (\omega) \bigg]
\label{eqn:Tmide}
\end{align}
where $T_{mid}$ is the mid-transit time, $P$ is the orbital period, $e$ is the eccentricity, and $\omega$ is the argument of periastron.

\begin{figure}[htb!]
\includegraphics[width=.4\textwidth]{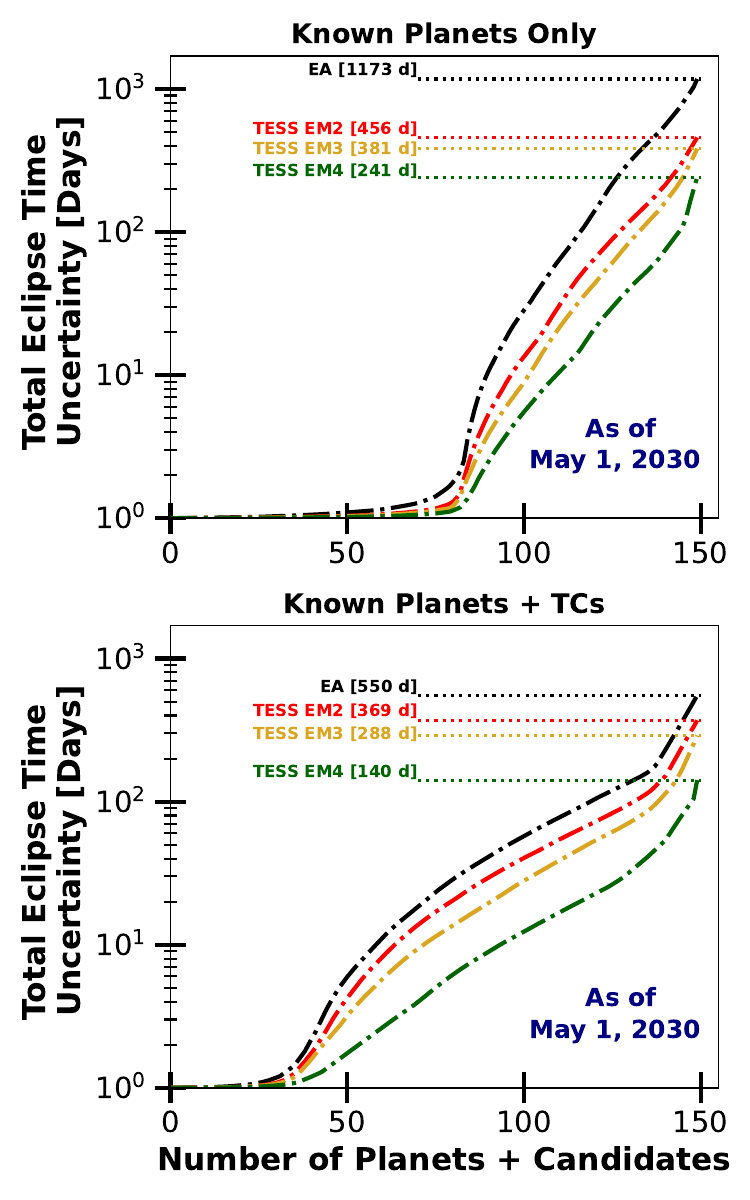}
\caption{The cumulative uncertainty for the mid-eclipse times of a future eclipse survey of our 150 object target lists, set to a nominal Ariel science start date of May 1 2030. For all planets and planet candidates, we assume an optimistic future uncertainty on $e\ \rm{cos}\ \omega$ of 10\%. In the worst-case scenario, where no ephemerides are refreshed (black lines) there will be over 500 of accumulated uncertainty for these targets by 2030. Observations by TESS's current extended mission (EM2, red lines) and our suggested extended missions (EM3 in gold, and EM4 in green) all serve to reduce these uncertainties and by EM4 the cumulative uncertainty falls below 240 days in both cases. Further decreasing the uncertainties for a 2030s eclipse survey will require targeted observing campaigns to either refresh the transit ephemeris close to the planned eclipse observation date or to capture one or more eclipses and determine the true eclipse epoch and spacing.}
\label{fig:Teclipse_Uncertainty}
\end{figure}

We implement a Monte Carlo method to calculate the cumulative mid-eclipse time uncertainties for the emission targets lists and set the transit ephemeris refresh dates to the planet's default transit epoch solution in the Exoplanet Archive and then to the last time the planet's transit is expected to be observed by the current TESS Extended Mission, and our proposed TESS Extended Missions 3 and 4 (Figure~\ref{fig:Teclipse_Uncertainty}). For simplicity, we assume that the 10-$\sigma$ mass requirement imposed on our emission targets translates to a 10\% uncertainty on the planets' $e\, \rm{cos}(\omega)$ values. 

Even with this high precision level and the relatively small eccentricities possessed by most of these targets (mean / median eccentricity = 0.037 / 0.0 for the known planets only list and  0.033 / 0.008 for the known planets + candidates lists) the total eclipse time uncertainty quickly becomes prohibitive. If no ephemerides are refreshed beyond what is posted in the Exoplanet Archive or in the TESS/Kepler planet candidate lists there will be 550 - 1173 days of uncertainty accumulated by 2030. Observations by TESS's current extended mission (EM2, red lines) and our suggested extended missions (EM3 in gold, and EM4 in green) decrease the total uncertainty from 550 days to 369, 288, and 140 days in the Known Planets + Candidates case and from 1173 days to 456, 381, and 241 days in the Known Planets Only case as Extended Missions 2, 3, and 4 are considered (Figure~\ref{fig:Teclipse_Uncertainty}). Yet both cases still results in more than 130 days of observing uncertainty in 2030.

Setting the epoch refresh date to include transits observed in our suggested EM4 for TESS decreases the uncertainty by a factor of 4.8 and 3.9 for the known planets only and known planets + candidates target lists, respectively (Figure~\ref{fig:Teclipse_Uncertainty}) but both cases still results in more than 130 days of observing uncertainty in 2030.

\section{Time Estimate to Obtain Necessary Radial Velocity Planet Masses}\label{sec:RVTime}

\begin{figure}[htb!]
\includegraphics[width=.45\textwidth]{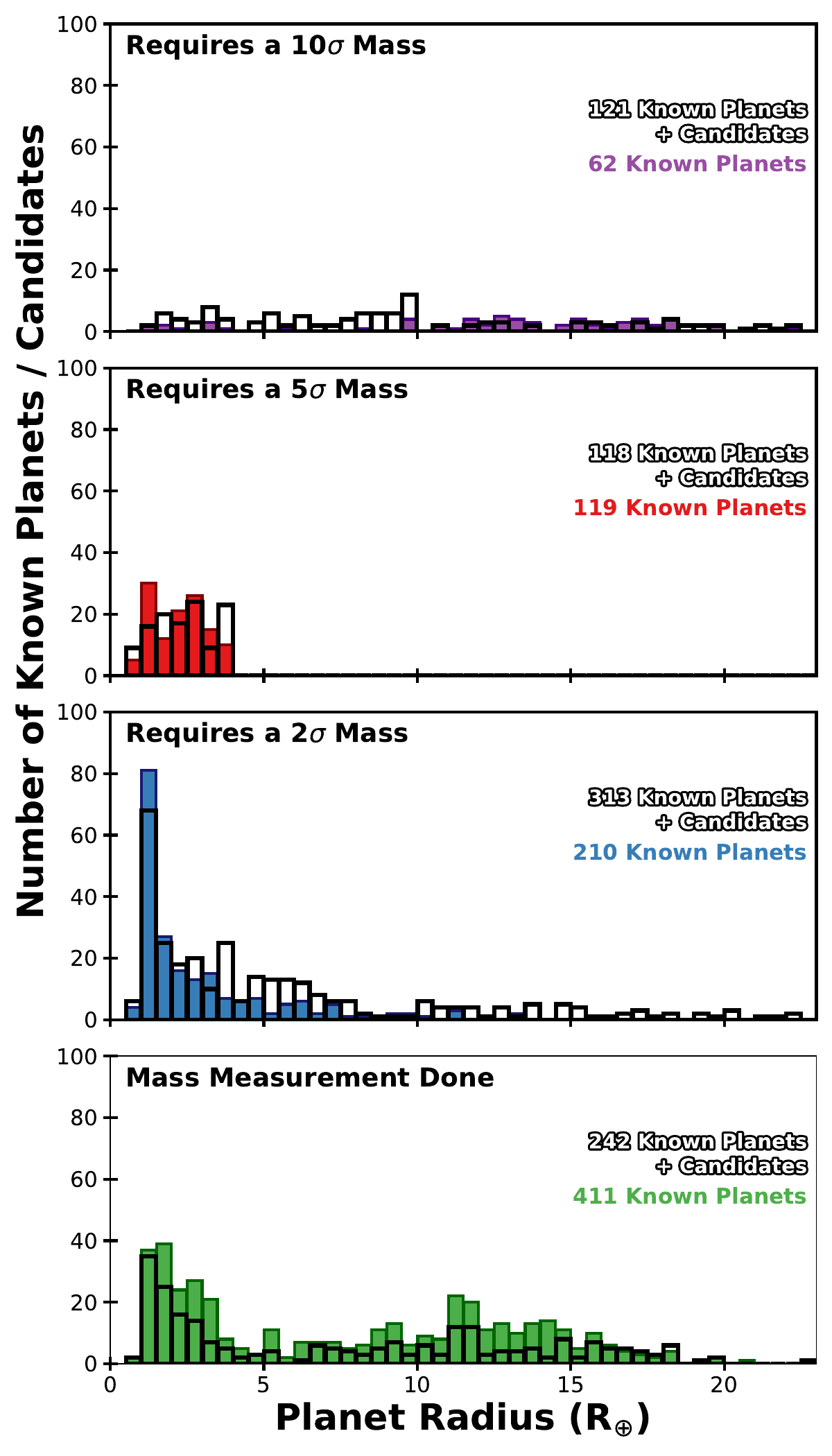}
\caption{Distribution of radii for the planets and planet candidates that still require a 10-$\sigma$ (top panel, reserved for eclipse targets), 5-$\sigma$ (second panel, for transmission targets that are both small \textit{and} cool), or 2-$\sigma$ (third panel, for transmission targets that are either large \textit{or} hot), and those that already have a sufficient mass measurement (bottom panel). Each panel includes a histogram for the mass needs distribution assuming the inclusion of both known planets and TESS/Kepler planet candidates (black outline histogram) and the use of only those planets that have already been confirmed (filled color histogram).}
\label{fig:MassNeedsHist}
\end{figure}

The targets that require a 10-$\sigma$, 5-$\sigma$, or 2-$\sigma$ mass precision, and those which already have a satisfactory mass precision in the literature, are summarized in Figure~\ref{fig:MassNeedsHist}. In both target prioritization schemes, the number of planets requiring mass measurements increases when moving from the known planets only target list to the known planets + candidates list -- unsurprising given that many exoplanet discovery publications include some attempt at quantifying the planets' mass. 
 
Exoplanet masses are generally measured via radial velocity (RV) observations of the host star which are taken with high resolution, ground-based spectrographs. The exact number of RV observations needed to achieve a specific mass precision for a given planet is challenging to predict. It depends on a variety of parameters that are often unknown at the onset of observing such as the star's RV variability due to phenomena such as spots and/or faculae \citep[e.g.][]{Aigrain2007,Meunier2019} and the presence of additional non-detected planets in the system. It also depends on more readily available values such as the RV information content of the star, the resolution of the RV spectrograph, and the total photon noise of the observation \citep{Beatty2015, Bouchy2001}. \citet{Cloutier2018} provides an analytical formalism to estimate the number of RV measurements of a given single measurement precision, determined by either white or correlated noise (due to, e.g., stellar activity), that are required to achieve a specified RV semi-amplitude uncertainty.

Equation~19 in \citet{Cloutier2018} asserts that\footnote{The use of 3 as the scalar value instead of the 2 seen in \citet{Cloutier2018} is the incorporation of the author's recommendation that the number of observations be scaled up by a factor of 1.5$\times$ in cases of moderate stellar variability. We adopt this scaling for all of the planets that require new or improved RV masses due to the lack of existing knowledge about their host stars' variability.}:

\begin{equation}\label{eqn:RV_Nobs}
    N_{RV} = 3 \bigg[\frac{\sigma_\mathrm{eff}}{\sigma_{\mathrm{RV}}}\bigg]^{2}
\end{equation}

\noindent where $\sigma_{\mathrm{eff}}$ is an \lq{}effective\rq{} RV uncertainty (composed of the quadrature sum of the RV noise floor of the spectrograph, the photon-noise limited RV precision, and any RV jitter arising from additional sources of RV noise such as stellar variability and unknown planets) and $\sigma_{\mathrm{RV}}$ is the uncertainty in the final RV semi-amplitude measurement. 

Converting from a number of observations to an amount of telescope time requires knowing the exposure time of each observation, which in turn requires specifications about what size of telescope and what kind of spectrograph will be used to obtain the RV measurements. To simulate a degree of realistic survey execution, here we consider the use of a mixture of medium (3.5-m) and large-scale (10-m) RV resources and make use of the Exposure Time Calculator (ETC) developed for the Keck Planet Finder \citep[KPF,][]{Gibson2016}, a recent extreme precision RV (EPRV) spectrograph deployed on the Keck I telescope. The exposure time calculator is based is upon a detailed radial velocity error budget \citep{Halverson2016} for the instrument. When provided with the effective temperature and V-magnitude of a star, the ETC produces an estimate of the RV photon noise expected from a user-specified exposure time. The results of the KPF exposure time calculator can be generalized to simulate similar instruments on different aperture facilities by scaling the output exposure time by the ratio of the collecting areas. Here we apply a scaling factor of 7.9, based upon the ratio of the effective collecting areas of the 10-m Keck I telescope and the 3.4-m WIYN telescope, which we use as an example of a 3.5-m class telescope.

\begin{figure*}[htb!]
\includegraphics[width=.95\textwidth]{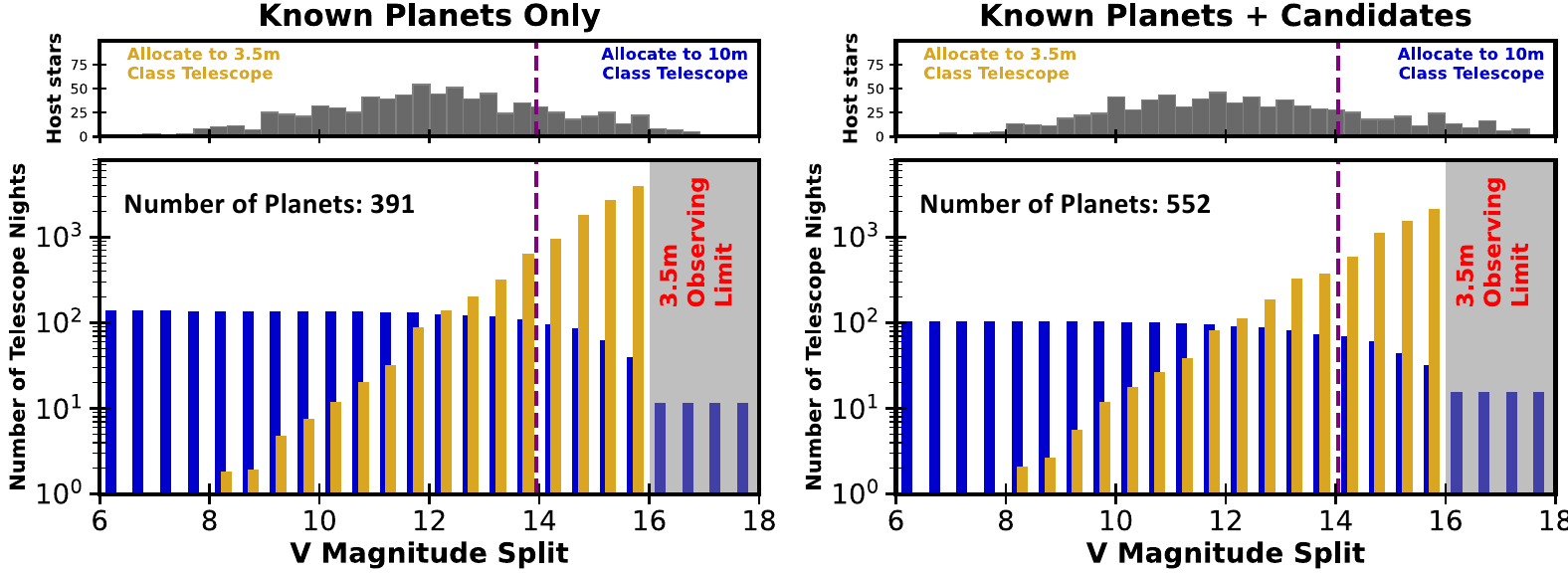}
\caption{Comparison of the number of telescope nights required on a 10-m class telescope (blue bars) and a 3.5-m class telescope (gold bars) to obtain sufficient mass measurements for the two representative target lists based on what V-magnitude cut is used to split targets between the two facilities. The purple dashed line in each plot shows the V-magnitude split where the total number of 3.5-m nights is $\sim$6x the number of 10-m nights, in accordance with the Cost $\propto$ Diameter$^{1.7}$ relation from \citet{Stahl2020}.}
\label{fig:RV_TimeSplit}
\end{figure*}

We first use the ETC and Equation \ref{eqn:RV_Nobs} to determine what RV precision an EPRV instrument on a 10-m telescope can produce for that planet's host star in a 30-minute exposure, as 30 minutes is generally used as an upper limit for a single RV exposure in order to avoid problematic levels of cosmic ray contamination. We then determine how many exposures at that RV precision are necessary to achieve the 2-, 5-, or 10$\sigma$ mass precision required for the planet in question. For planets where the number of required observations is less than eight, we instead set N$_{\mathrm{obs}}$ = 8 and then rearrange Equation \ref{eqn:RV_Nobs} to determine the individual RV measurement precision necessary to reach the required planet mass precision in eight observations. Requiring a minimum of 8 observations allows a well executed survey to ensure that the resulting RV measurements can be spread throughout the planet's orbital phase curve to allow measurement of the orbital eccentricity and longitude of periastron \citep{Burt2018}. When tabulating the total telescope time needed, we account for an average overhead time (telescope slew and acquisition) of 120 seconds for the 10-m class telescope and 180 seconds for 3.5-m class telescope, adopted from \citet{Chontos2022} and \citet{Gupta2021}, respectively.
 
When deciding which objects to assign to the 10-m class facility and which to assign to the 3.5-m class facility, we split the targets using a host star magnitude cut. All planets orbiting stars brighter than the magnitude cut are followed up on the 3.5-m class telescope while all planets orbiting stars fainter than the magnitude cut are followed up with the 10-m class telescope. To select this limiting magnitude, we consider the rough cost ratio of the two facilities based upon the parametric cost model for ground-based telescope presented in \citet{Stahl2020}, which finds that optical telescope assembly costs scale with telescope diameter as: $\mathrm{OTA\$} \; \propto \; D^{1.7}$. We note that the cost ratio described above encapsulates the one time assembly cost of such facilities, and does not address ongoing operational costs, but it serves as a useful metric for our simulated survey.

Inserting the sizes of our representative telescopes, 10-m and 3.5-m, produces a cost ratio of $\sim$6, and so we identify a V-magnitude cut in each of the target lists that results in roughly six times more nights on the medium class telescope than on the large telescope. Figure \ref{fig:RV_TimeSplit} and Table \ref{tab:RV_TelTime} show how the amount of time required on each telescope varies with the V-magnitude cut used in each of the four target selection schemes. We also include the number of nights it would take to obtain all of the necessary planet masses using just the 10-m telescope in Table \ref{tab:RV_TelTime} to allow for more direct comparisons between the target lists.

\begin{table*}
  \centering
  \renewcommand{\arraystretch}{1.2}
  \begin{tabular}{|l|c|c||c|c|}
    \hline
     & \multicolumn{2}{c|}{\textbf{Known Planets Only}} & \multicolumn{2}{c|}{\textbf{Known Planets + Candidates}} \\
    \hline
    \hline
     & 10-m Only & 10-m + 3.5-m & 10-m Only & 10-m + 3.5-m \\
     & & V-mag Split: 13.95 & & V-mag Split: 14.05 \\ 
     \hline
    10-m Planets & 391 & 120 & 552 & 152 \\ 
    \hline
    3.5-m Planets & 0 & 271 & 0 & 400 \\ 
    \hline\hline
    10-m Nights & 138 & 99 & 105 & 69 \\ 
    \hline
    3.5-m Nights & 0 & 581 & 0 & 415 \\ 
    \hline      
  \end{tabular}
  \caption{Summary of the estimated telescope time required to obtain masses of the necessary 2-, 5-, or 10-$\sigma$ precision for the representative planet populations defined above. The table lists the time necessary if the entire RV follow-up program were carried out on a 10-m class telescope and the time needed if the targets are instead split between a 10-m and a 3.5-m class telescope using a V-magnitude cut that produces a telescope allocation ratio of roughly 6x as per the 10-m:3.5-m cost ratio derived using \citet{Stahl2020}.}\label{tab:RV_TelTime}
\end{table*}

The least RV resource intensive option is to select targets from the combined list of known planets and TESS/Kepler planet candidates. When considering a split 10-m + 3.5-m telescope approach and requiring a factor of 6 increase from the number of 10-m nights to the number of 3.5-m nights as per \citep{Stahl2020}, this target list would require 69 nights on a 10-m class telescope and 415 nights on a 3.5-m class telescope. If the same survey were to be carried out using only the 10-m telescope, the required number of 10-m nights would increase from 69 to 105. If we instead restricted the target list to only include those planets that have already been confirmed in the scientific literature, as is often a requirement for space-based spectroscopic follow up (e.g., on HST and JWST review panels to avoid investing time on targets that are not real planets) then the RV telescope requirements increase to either 99 nights of 10-m + 581 nights of 3.5-m time or 138 nights of 10-m telescope time alone. This is true even though the total number of planets that require mass measurements decreases when moving from the known planets + candidates case to the known planets only case -- from 552 objects down to 391.

\begin{figure}[htb!]
\includegraphics[width=.45\textwidth]{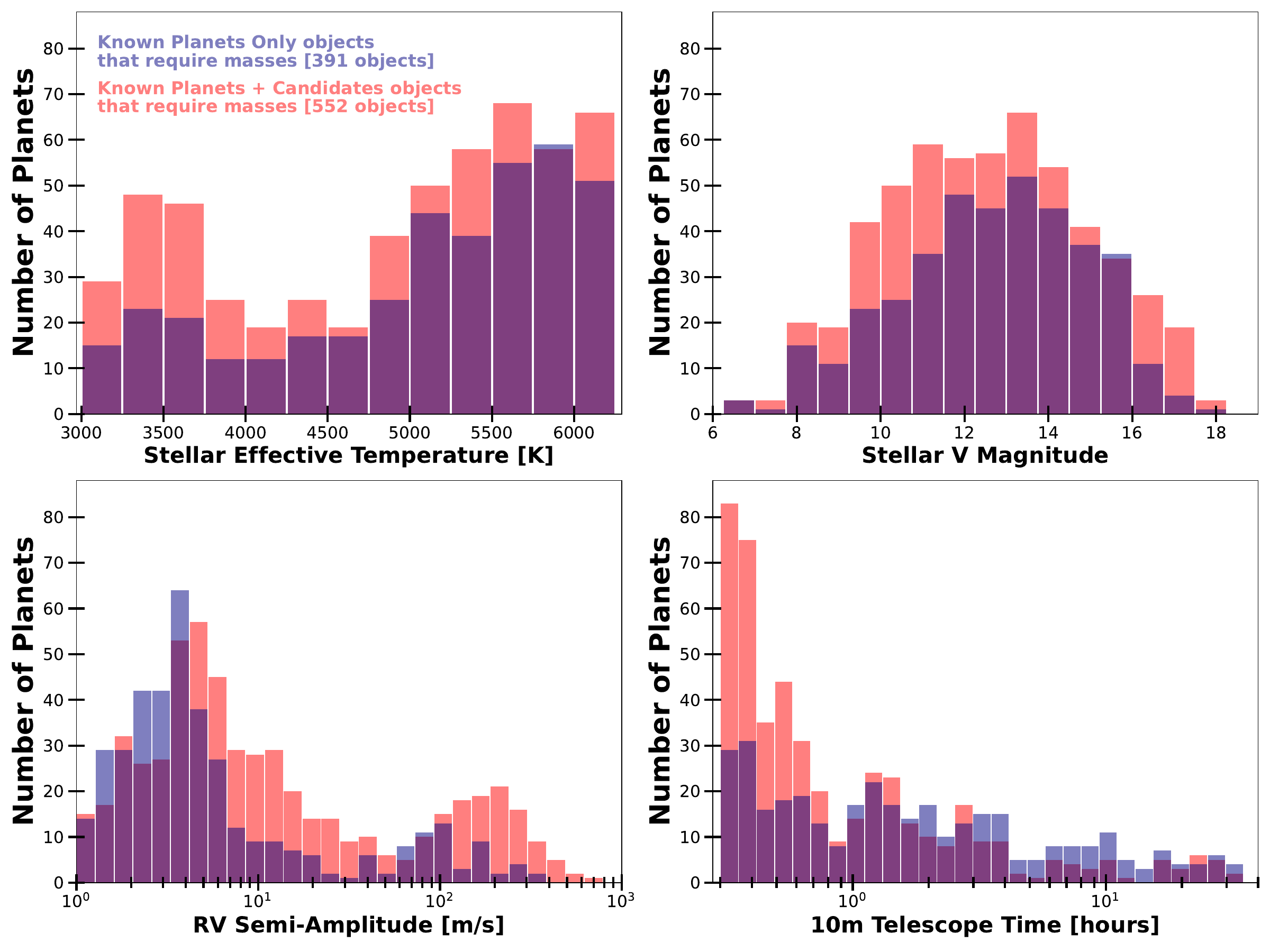}
\caption{Comparison of select stellar and planet properties for the objects that require RV follow up efforts from the known planets only target list (red) and the known planets + candidates target list (blue). While the known planets + candidates list contains 41\% more targets that require mass measurements, a combination of those targets orbiting brighter stars, cooler (and thus lower mass) stars, and having larger RV semi-amplitudes results in the known planets + candidates RV survey still requiring less telescope time than the known planets only RV survey.}
\label{fig:RV_TargetComparison}
\end{figure}

This seeming discrepancy, where the list with more objects that require masses actually takes less telescope time to complete, is driven by a combination of the host star and planet properties (Figure \ref{fig:RV_TargetComparison}). The known planets + candidates RV target list features brighter host stars and more M-dwarf host stars, both of which lead to shorter RV exposure times when all else is held equal thanks to the increase in photons from the brighter stars and the increased RV information content from the cooler host star \citep{Beatty2015}. The RV semi-amplitudes of the known planets + candidates objects also skew larger than the known planets only objects. All of these features combine to produce individual telescope time estimates that are lower for the known planets + candidates objects than for the known planets only objects, resulting in a 30\% decrease in the total 10-m telescope time even though the RV target list is 41\% larger.

\section{Discussion and Results}\label{sec:discussion}

\subsection{Ephemeris Refinement Campaign Recommendations}\label{sec:ephemeris_refinement_recommendations}
While TESS is naturally performing ephemeris maintenance \citep[e.g.,][]{dragomir20}, its full sky observing approach means that most areas of the sky are revisited only every two or more years. It is therefore likely that many transits for any individual planet will be missed by the spacecraft. This issue is exacerbated further for longer period, more temperate planets that transit their stars less frequently, where key transits are more likely to be missed. In addition, some dimmer and/or shallower transits could prove difficult for TESS to perform ephemeris maintenance on without multiple observations. A third extended mission has already been proposed for TESS and ESA's Plato mission \citep{rauer14} will also have the capability for ephemeris refinement via its transit observations. But it would not be prudent to rely upon these two spacecraft alone to perform the entirety of this critical ephemeris maintenance. Fortunately targeted ground-based efforts to routinely observe known transits can play a crucial role in retiring the risk of missed, or overly expensive, atmospheric spectroscopy observations.

To determine which targets will require ephemeris maintenance to limit future atmospheric characterization overheads we calculate which targets' ephemerides become \lq{}stale\rq{}, wherein the 1$\sigma$ uncertainty on the ephemeris exceeds half the transit duration \citep{zellem20}, by the 1 July 2025 start of JWST Cycle 4 and our assumed 1 May 2030 start date of Ariel's science operations. We find that for both of our target lists at least 150 objects require ephemeris maintenance ahead of 2025 and over 180 will require ephemeris updates ahead of 2030 (Table~\ref{tab:stale_numbers}). We assume a campaign design where each observation requires as much out-of-transit data as in-transit data which sets the observing time requirement to twice the transit duration. We also note that due to scheduling/location constraints a single ground-based facility can generally observe only 1--2 transits in a single night. Thus, even in the best of circumstances (no weather losses, no instrument/facility down time, and having a visible transit event of a target that requires ephemeris refinement on any given night) this ephemeris maintenance campaign would take a single observatory over 100 full nights to conduct. Planets orbiting dimmer stars and those with shallower transits will result in less precise measurements and may require multiple observations for ephemeris maintenance, so these estimates are a lower bound.

\begin{table*}\label{tab:stale_numbers}
\centering
\begin{tabular}{l | c | c }
\hline
\hline
\textbf{}  & \textbf{Targets Requiring} & \textbf{Total Photometry}  \\
\textbf{Observation Date}  & \textbf{Ephemeris Maintenance} & \textbf{Hours Required} \\
\hline
July 1 2025: KP Only & 152 & 900 \\
July 1 2025: KP + PCs & 154 & 644 \\
\hline
May 1 2030: KP Only & 187 & 1106 \\
May 1 2030: KP + PCs & 199 & 807 \\
\hline
  \end{tabular}
  \caption{Ephemeris maintenance required prior to JWST Cycle 3 and Ariel's predicted science start in order to avoid ``stale" transit ephemerides, based on the transit epochs reported in the Exoplanet Archive (for known planets) and the TEV list (for TESS planet candidates). Stale is defined as the 1$\sigma$ uncertainty on the ephemeris exceeding half the transit duration by the assumed observation dates: July 1, 2025 for JWST Cycle 4 and May 1, 2030 for Ariel.}
\end{table*}

Given the all-sky nature of the representative target lists and the $\geq$100~nights required to carry out sufficient ephemeris maintenance (Table~\ref{tab:stale_numbers}), a large, globally-distributed network of telescopes is an ideal resource to ensure full-sky coverage of both Northern and Southern Hemispheres and to mitigate impacts due to weather and night time observing constraints. Two such programs that have been established to leverage community (\lq{}citizen\rq{}) scientists to aid in ephemeris maintenance are the US-based Exoplanet Watch\footnote{https://exoplanets.nasa.gov/exoplanet-watch/} \citep{zellem20, pearson22} and the European-based ExoClock\footnote{https://www.exoclock.space/} \citep{kokori22, kokori23}. The former has been established to support all NASA efforts (e.g., observations with HST, JWST, and large ground-based telescopes) while the latter has been formed by the Ariel mission to focus explicitly on the Ariel target list. Both of these projects are already performing ephemeris refinement at scale.

When considering eclipse observations, refinement of the eclipse ephemerides by directly observing eclipses will prove difficult to execute at scale from the ground given their smaller signal size. A typical eclipse signal for a hot Jupiter around a solar-type star is only $\sim$0.1\% and when examining the eclipse spectroscopy targets on our target lists we find that only 7-8 of the 150 planets on each emission spectroscopy target list have a predicted eclipse depth \textgreater\ 0.1\%. 

Observing such small signals with ground-based telescopes and spectrographs has proven difficult and only a handful of potential detections \citep[e.g.,][]{swain10, thatte10, waldmann12, zellem14b} have been conducted from the ground. The data analysis can also be difficult, requiring novel reduction techniques, which have been questioned in the field \citep[e.g.,][]{mandell11}. While ground-based photometric observations of eclipses are possible --- for example, \citealt{ORourke2014} have successfully detected signals down to 0.1\% in K$_{\rm{S}}$ band using the 5-m Palomar telescope --- performing a cohesive eclipse ephemeris maintenance program of 150 targets with the $\geq$5-meter class telescope(s) required for such work would be difficult, both in terms of scheduling and in achieving the high precision necessary to detect an eclipse. Therefore, we instead recommend that the maintenance of eclipse times be addressed by refining the mid-transit times via photometry while the orbital eccentricity and longitude of periastron are determined via RV follow up.

As transit light curves do not provide constraints on the eccentricity and longitude of periastron, these values must be determined from a Keplerian model fit to RV data points. We therefore advocate for precisely predicting upcoming eclipse events via a joint analysis of transit and RV data \citep[e.g.,][]{pearson22}. Transit photometry can be conducted with relatively small telescopes \citep[e.g.,][]{zellem20} to calculate the impact parameter (to predict if the planet will actually be eclipsed by its host star), transit ephemeris, and orbital period while RV data from larger facilities can constrain the orbital eccentricity and argument of periastron. Taken together these can provide strong constraints on the next estimated eclipse time.

\subsection{Overall RV Resource Requirements}

Knowing that hundreds of nights of 10-m class telescope time along with many hundreds of nights of 3.5-m class telescope time are necessary to obtain the recommended mass measurements to enable the accurate characterization of exoplanet atmospheres suggests the need for a large, sustained, and collaborative mass measurement effort. Even the Keck Strategic Mission Science (KSMS) program, which is intended to directly support NASA mission science goals, can only allocate 15 nights of telescope time spread between Keck I and Keck II per semester\footnote{https://nexsci.caltech.edu/missions/KeckSolicitation/ksms.shtml}. 
If we assume that the generation of a well vetted atmospheric characterization target list is prioritized enough to claim half of the KSMS time allocation, our known planets + planet candidates target list (the less resource intensive of the two representative target lists) would take 9 semesters to execute on Keck \textit{in addition to} a significantly longer companion survey on a 3.5-m class telescope (e.g. using the NEID spectrograph at the WIYN telescope) that would require 415 nights of telescope time.

While our simple two telescope model highlights one way to distribute the observing burden, a more realistic RV campaign would span a wide range of facilities around the globe. Comprehensive northern and southern coverage will be needed thanks to the on-sky distribution of the target stars (Figure~\ref{fig:SkyMap}) and there is ample opportunity for RV spectrographs on telescopes as small as 0.7-m to contribute to this effort due to the wide range of RV semi-amplitudes (1--600 m s$^{-1}$) and host star magnitudes included in the target lists (Figure~\ref{fig:RV_TargetComparison}). A coordinated effort where targets are spread among telescopes and spectrographs of varying sizes and precision floors would lead to a more efficient and effective mass measurement campaign. The enormous scope of these efforts highlights the need to continue supporting confirmation work for the planets candidates alongside work to refine the masses and orbital ephemerides of the confirmed planets as soon as possible so that the survey neither overwhelms the available RV telescope resources nor fails to finish in time for missions like Ariel.

\subsection{RV Target Prioritization Within the Target List}

A second crucial decision when planning future RV characterization surveys concerns the prioritization of the planets that require new or better mass measurements within a specified target list. Starting from the representative ranking target list that includes both known planets and planet candidates, we can rank the planets that require masses in a variety of ways:
\begin{enumerate}
    \item FOM Ranking: Start by following up the planet with the highest FOM$\mathrm{_{transit}}$, and proceed down the list until reaching the planet with the lowest figure of merit value. This method ensures that the planets / candidates most likely to produce high quality transmission spectroscopy observations obtain mass measurements first.
    \item Efficiency Ranking: Order the planets by how much RV time they require, and begin with the planets that require the least time. This method ensures that the largest number of planets obtain their requisite mass measurements in the shortest amount of time by relegating the most time-intensive planets to the end of the survey.
    \item Homogeneous Spread Ranking: Similar to the methodology used to construct our representative target lists, this system moves iteratively through the thirty \Teq\ and \Rearth\ bins described in Section \ref{sec:rep_trans} and in each iteration selects the planet with the lowest RV time requirement in each bin. This method prioritizes the equal representation of planets across the \Teq\ and \Rearth\ axes in the least amount of RV telescope time possible.
\end{enumerate}

Any of these ranking systems could be justifiable choices when executing an RV survey in support of current or future transiting exoplanet science cases. Each approach, however, produces different combinations of the total number of planets and the types of planets that obtain their necessary mass measurements within a given number of telescope nights. 

\begin{figure*}[htb!]
\includegraphics[width=.9\textwidth]{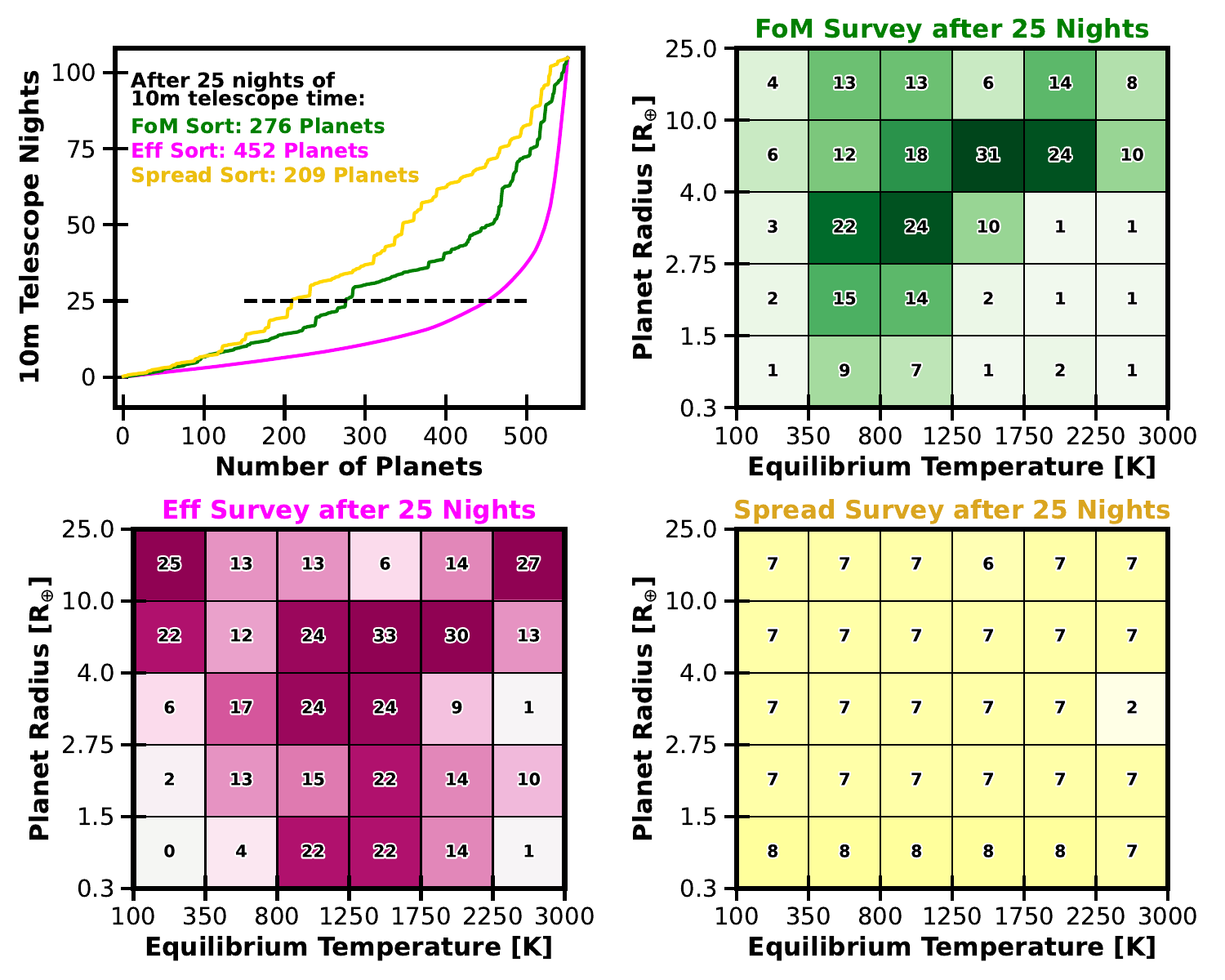}
\caption{Comparison of the overall planet mass acquisition rate (top left) and mid-survey populations of planet masses using three different target prioritization schemes: a Figure of Merit ranking (green line and top right grid), an RV Efficient ranking (pink line, and bottom left grid), and a Homogeneous Spread ranking (gold line, and bottom right grid). Upon completion of the full 105 night survey on Keck, all survey designs converge and provide requisite masses for all planets on the target list (top left). But the order in which planet masses are acquired varies and after 25 nights the populations of planets with newly measured masses within each survey (top right and bottom row) are quite different. Selecting a survey design that enables, for example, the early mission science priorities of the Ariel science team can ensure high quality science even if the full RV survey is not completed before Ariel begins returning data.}
\label{fig:RV_25night_PlanetDist}
\end{figure*}

For simplicity's sake, we assume that all RV follow-up efforts are carried out using a 10-m class facility which would require 105 nights of observing time to obtain masses of sufficient precision for the 552 planets and candidates on the Representative Ranking target list that require new or improved mass measurements (Table~\ref{tab:RV_TelTime}). By the end of those 105 nights, all ranking systems converge to having obtained masses for the entire set of planets. But if we examine an intermediate point of the survey, such as 25 nights into the total program, the impact of each prioritization becomes more apparent (Figure \ref{fig:RV_25night_PlanetDist}). After 25 nights of telescope time, the `Efficient' survey has already obtained masses for 452 of the 552 planets in need, whereas the FOM survey and the Homogeneous Spread survey have only provided masses for 276 and 209 planets, respectively. 

Both the `Efficiency' and `Homogeneous' survey approaches provide numerous planets with the requisite mass precision across the majority of \Rearth\ and \Teq\ bins. They therefore allow for comparative planetology to happen within most regions of exoplanet parameter space much earlier in a space-based atmospheric characterization survey than the `FOM' RV survey design would. And in most bins, the `Efficiency' approach delivers 2-4 times more planets than the `Homogeneous' approach, which furthers the potential for examining the impact of exoplanet characteristics beyond just the radius and equilibrium temperature. So for broad science applicability the `Efficiency' survey provides the largest planet population to choose from for early atmospheric characterization efforts.

The few exceptions to this result are found mostly among the smallest and coolest planet bins, unsurprising as a planet's RV semi-amplitude is proportional to its mass and inversely proportional to its orbital period. Smaller (generally lower mass) and/or cooler (generally longer period) planets induce smaller RV semi-amplitudes than their larger and/or hotter counter parts and require more RV telescope time to acquire a sufficiently precise mass measurement. That increased telescope time requirement prevents these planets from being addressed in the first 25 days of the `Efficiency' survey, and so these bins remain unpopulated. We note that the smallest and hottest planet bin (R$_{p}$ = 0.3 - 1.5 \Rearth\ and \Teq = 2250 - 3000 K) is also underpopulated at the 25 day mark in the `Efficiency' survey compared to the `Homogeneous' survey. In this case, the lack of planets delivered by the `Efficiency' survey is because the host stars in that bin are notably fainter (average Vmag = 13.9) than those in the neighboring, lower \Teff, bin (average Vmag = 12.1) which again drives up the RV time requirement. This serves as a useful reminder that two axes of consideration are insufficient to capture the full variety of both planet and host star parameters present in a population of hundreds of exoplanets.

\subsection{Additional Benefits of Using Photometric Data to Inform RV Mass Measurement Surveys}

Following up transiting planet candidates, rather than pursuing more traditional, uninformed RV surveys that target stars without a priori knowledge of their planetary systems simplifies the mass measurement process in a number of ways. Having a precise period and transit epoch from the photometric data provides narrow priors for two parameters of the radial velocity orbital solution, decreasing the number of RV data points required to obtain a precise mass \citep{Cloutier2018}. It also provides the ability to make informed choices of where to place observations along the planet's orbital phase curve which contributes to more efficient RV mass measurements and also to more precise measurements of the planet's eccentricity \citep{Burt2018, Gupta2024, Lam2024}. Using the transit data to determine the planet's orbital inclination with respect to its host star also removes the sin(i) degeneracy that RV orbital fits alone are subject to, which in turn removes an (on average) factor of two uncertainty in the planet's true mass.

If the host star exhibits rotational modulation due to spots or faculae, photometric data can help to constrain the stellar rotation period, determine whether the star is single or double spotted, and provide a sense of the spot evolution timescale on a particular star \citep[e.g.][]{henry99, zellem10,Kosiarek2020}. These three characteristics are represented in the hyperparameters of the quasi-periodic Gaussian Process kernel which (among other kernels) is now commonly used to disentangle radial velocity signals from a star from the Keplerian signal induced by an orbiting planet \citep[e.g.][]{Rajpaul2021,Barragan2022,Nicholson2022}. Thus, ground-based photometry is not only necessary to improve ephemeris refinements, but can can also provide key support in both the planning of RV surveys and in the interpretation of the resulting data.

Lastly, while we focus on the value of ground-based transit photometry (to provide updated ephemerides) and radial velocity (to provide planetary masses and updated orbital parameters), these two methods can be combined to provide even more accurate orbital parameters \citep[e.g.,][]{pearson22}. For example, when transits and radial velocity measurements are jointly analyzed, they can provide significant ($\sim10\times$) improvements to the orbital and planetary parameters versus analyzing the transit or radial velocity measurements alone \citep{fruber23}.

\section{Conclusions}\label{sec:conclusions}

With the operations of NASA's HST and JWST, and the pending launch of NASA's Pandora and Roman missions and ESA's Ariel Mission, the exoplanet scientific community is entering an era of unparalleled opportunities for comparative exoplanetology. The study presented here explores the creation of a more representative atmospheric characterization survey target list that draws planets as evenly as possible across a range of planet radii and equilibrium temperatures, but still prioritizes the planets with the most promising transmission and emission Figure of Merit value within each \Teq\ and \Rearth\ grid space. We generate two such target lists, one that considers only known confirmed planets and one that draws from the combined set of confirmed planets and TESS/Kepler planet candidates. We further present an assessment of the current state of orbit and mass determination for confirmed planets and TESS/Kepler candidates as listed at the NASA Exoplanet Archive and the TEV website.  

We note that the recommendations we make here regarding target list assembly are designed specifically to maximize the uniform sampling of the exoplanet \Teq\ vs \Rearth\ phase space. Designing exoplanet samples optimized for specific science questions will likely lead to other ranking metrics and target selection recommendations, and observers should consider carefully which planet and/or host star parameters will have the most impact on the science question that they are attempting to answer. As a starting point, such efforts could extend this grid into 3rd or 4th dimensions to incorporate additional planetary and stellar parameters in the selection process. Such investigations and recommendations for alternative target list assembly approaches are a promising topic for future work.

Spectroscopic observations of exoplanet atmospheres require sufficiently precise planetary masses in order to accurately and precisely determine their chemical abundances.  Nearly as important as the need for a quality planetary mass for scientific interpretation is the need for quality orbital parameters, which set the observing efficiency of the survey. Our understanding of exoplanet atmospheres relies heavily on transit and eclipse spectroscopy, and knowing when to observe is critical to making the most effective and efficient use of telescope resources.

We find that 20+\% of the targets on our example survey lists require some level of transit ephemeris maintenance before observation by future missions such as Ariel in the early 2030s. Without this maintenance the cumulative uncertainty on the times of transit and eclipse can grow to 2-3 years, exceeding half the prime mission lifetime of Ariel. The TESS mission provides a valuable source of ephemeris updates, especially if future Extended Missions are approved and continue an all sky observing approach. Yet the same all sky survey design that allows TESS to capture the transits of thousands of planets and planet candidates also results in roughly two year observing gaps across the majority of the sky. Targeted ground-based observing efforts will therefore be needed to fill in these gaps and provide updated transit ephemerides for key objects as planned space-based observations draw near.

For the planetary masses we find that the fraction of targets needing accurate masses depends strongly on whether planet candidates are considered alongside previously confirmed planets in the target selection process. When considering only known planets, $\sim50\%$ of the target list still requires a mass measurement of sufficient precision. That number increases to $\sim70\%$ when the list draws from both known planets and planet candidates. This study estimates the amount of RV facility telescope time that would be needed to obtain masses for each of the target scenarios and finds that the requirements range from one hundred to many hundreds of nights spread across a combination of large ($\sim$10-m) and mid-sized ($\sim$3-m) telescopes. Yet the total telescope time required is lower in the known planets + candidates case as that list features brighter and cooler host stars and objects that induce larger RV semi-amplitudes, all of which contribute to a smaller amount of telescope time needed to obtain a specific mass precision.

Out of this study, we have identified a series of recommendations that could make community efforts for preparation and interpretation of atmospheric characterization endeavors more effective and efficient in advance of transit observations.  

\begin{itemize}
    \item A continued and coordinated effort to confirm the planetary nature of the ever increasing number of TESS planet candidates, many of which may become high priority targets for observing with JWST and Ariel.

    \item Regular updates to the transit ephemerides of both known planets and planet candidates that incorporate all available TESS data.

    \item Continued monitoring of transiting planets with Exoplanet Watch, ExoClock, TESS, and (soon-to-be-launched) PLATO, and the determination of accurate and precise transit ephemerides across all of the transit observations. 
    
    \item A continued and well coordinated effort to obtain sufficiently precise and accurate masses and orbits for exoplanets expected to be the best targets for observing with HST, JWST, Pandora, and Ariel. 

    \item A more consistent treatment and full reporting of orbital, transit, and eclipse parameters in the published literature by the community to enable services like the NASA Exoplanet Archive, ExoFOP, Exoplanet Watch, and ExoClock to have a more reliable and understandable set of parameters from which the community and the missions can draw.

\end{itemize}

\section*{Acknowledgments}
We follow the guidelines \citep{10.5281/zenodo.10161527} of NASA’s Transform to OPen Science (TOPS) mission  for our open science practices. The software used to execute most analyses in this paper and to generate the corresponding figures is hosted on GitHub \footnote{https://github.com/JenniferBurt/exoatmosphere-targetlists} and is preserved, alongside the input data sets, on Zenodo at \dataset[doi:10.5281/zenodo.15660705]{https://doi.org/10.5281/zenodo.15660705}.

Part of the research was carried out at the Jet Propulsion Laboratory, California Institute of Technology, under contract with the National Aeronautics and Space Administration. Copyright 2025. All rights reserved.

This research has made use of the NASA Exoplanet Archive and the Exoplanet Follow-up Observation Program (ExoFOP) \citep{10.26134/ExoFOP5} website, which are operated by the California Institute of Technology, under contract with the National Aeronautics and Space Administration under the Exoplanet Exploration Program.

This publication makes use of data products from Exoplanet Watch, a citizen science project managed by NASA's Jet Propulsion Laboratory on behalf of NASA's Universe of Learning. This work is supported by NASA under award number NNX16AC65A to the Space Telescope Science Institute, in partnership with Caltech/IPAC, Center for Astrophysics|Harvard \& Smithsonian, and NASA Jet Propulsion Laboratory.

We thank the team behind the board game \textit{Forbidden Island} for helping us visualize and describe how we select our ``Representative Sample'' of planets.

\software{
\texttt{astropy} \citep{astropy2013, astropy2018, astropy2022},
\texttt{MRExo} \citep{Kanodia2019, Kanodia2023},
\texttt{TESS-point} \citep{tesspoint}
}
\clearpage
\bibliography{main}{}
\bibliographystyle{aasjournal}

\end{document}